# Performance vs Persistence : Assess the alpha to identify outperformers

Rémi Genet
Université Paris Dauphine
remi.genet@dauphine.eu

Hugo Inzirillo
Université Paris Dauphine
hugo.inzirillo@dauphine.eu

**Abstract**

*The number of pension funds has multiplied exponentially over the last decade. Active portfolio management requires a precise analysis of the performance drivers. Several risk and performance attribution metrics have been developed since the 70s to guide investors in their investment choices. Based on the study made by Fama and French (2010) we reproduce the experiment they had carried out in order to complete their work using additionnal features. Throughout this study we draw a parallel between the results obtained by Fama and French (2010) with the 3-factor model. The aim of this paper is to assess the usefulness of two additional factors in the analysis of the persistence of alphas. We also look at the quality of the manager through his investment choices in order to generate alpha considering the environment in which he operates.*

# 1 Introduction

**Mutual Funds**

Historically, mutual funds originated in the United States during the 19th century, and as the main operators in the US markets, mutual funds have continued to develop over the years. The ICI [1] states that the volume of assets held by mutual funds has risen from 1 trillion to 7 trillion in the space of ten years Reid (2000) . This rise accelerated in the early 2000s with the arrival of innovative management models, including the factor models that gave rise to smart beta strategies but also to Exchange Traded Funds.

Mutual funds, however, do not have unanimous support. Indeed, studies of Fama and French (2010) or Barras et al. (2010), come to the conclusion that they could not offer an over-performance exceeding their costs.

**Multiple factor models**

In the 60s, a first single-factor model was introduced following the work of Markowitz (1952). This financial asset pricing model (CAPM) emphasises market equilibrium driven by supply and demand for each asset. The purpose of this single-factor model is to measure the expected return on a risky asset by estimating the systematic risk of the asset $i$ noted $\beta^i$ .

$$\mathbb{E}[R_i] \; = R_f \; + \; \beta^i(\mathbb{E}[R_m] - R_f) \tag{1}$$

$\mathbb{E}[R_i]$ denote the expected return of the i-th asset. $\mathbb{E}[R_m]$ denote the expected market return and $R_f$ the risk free rate. The theoretical basis and unrealistic assumptions of the CAPM model have led to an interest among researchers in developing derived models. However, the CAPM remains one of the most widely used models in the financial industry. Following the numerous criticisms of the CAPM, many researchers have focused on linking asset returns to possible explanatory factors other than systematic risk that might explain these returns. Multi-factor models aim to include several sources of risk that explain expected returns with more details. Fama and French (1993), proposed a three-factor model to explain anomalies previously unexplained by the CAPM.

$$\mathrm{R_{i,t} - R_{f_t}} = \beta_0^{\mathrm{i}} + \beta_1^{\mathrm{i}}(\mathrm{RM_t - Rf_t}) + \beta_2^{\mathrm{i}}\mathrm{SMB_t} + \beta_3^{\mathrm{i}}\mathrm{HML_t} + \epsilon_{\mathrm{i,t}} \tag{2}$$

The market risk premium $R_{M_t} \; - \; Rf_t$ is the excess return induced par higher risk. The $SMB$ and $HML$ are two additional factors. SMB denote the difference between expected return of Small-Cap Stocks and Large-Cap Stocks. The HML factor refers to the difference between Growth-Stocks and Value-Stocks.

This three-factor model provides a concrete response to the CAPM anomalies by adding two additional factors: the SMB and HML, which explain the return on market capitalization and the book to market ratio, respectively. The empirical study showed that the size factor (SMB) had a strong impact on explaining returns. Smaller funds would be inclined to have higher expected returns than larger funds given their higher risk. The study also found that funds with a lower undervalued book to market ratio appeared to have higher returns than those with a higher overvalued ratio.

This extension of the Financial Asset Pricing Model (CAPM) has guided academic research. Extensions and derivations of the three-factor model have emerged. Carhart (1997) introduced a fourth factor in its explanatory model of mutual fund returns. In his work, Carhart adds an additional component to Fama and French's three-factor model, which is the momentum noted MOM.

$$\mathrm{R_{i,t} - R_{f_t}} = \beta_0^{\mathrm{i}} + \beta_1^{\mathrm{i}}(\mathrm{RM_t - Rf_t}) + \beta_2^{\mathrm{i}}\mathrm{SMB_t} + \beta_3^{\mathrm{i}}\mathrm{HML_t} + \beta_4^{\mathrm{i}}\mathrm{MOM_t} + \epsilon_{\mathrm{i,t}} \tag{3}$$

The additional factor in this model noted MOM is the difference between the expected returns on Low Momentum stocks and High Momentum stocks. This factor highlights the Momentum effect on expected returns and provides an additional explanation for a fund's returns. Momentum measures the change in the price of an asset over a given period of time. Fama and French (2015) add two new factors to their initial model: the RMW and the CMA as follow :

$$\mathrm{R_{i,t} - R_{f_t}} = \beta_0^{\mathrm{i}} + \beta_1^{\mathrm{i}}(\mathrm{RM_t - Rf_t}) + \beta_2^{\mathrm{i}}\mathrm{SMB_t} + \beta_3^{\mathrm{i}}\mathrm{HML_t} + \beta_4^{\mathrm{i}}\mathrm{RMW_t} + \beta_5^{\mathrm{i}}\mathrm{CMA_t} + \epsilon_{\mathrm{i,t}} \tag{4}$$

The addition of these two factors contributes significantly to the yield decomposition. The RMW measures the difference between the profitability of companies with a high and a low operating profit before interest and taxes. The CMA, on the other hand, measures the difference between the profitability of companies that invest with a conservative profile and those that invest with an aggressive profile.

[1] Investement Company Institue available at : https://www.ici.org/

**Alpha persistance**

This paper has two purposes. Starting from Fama and French (2010) we wanted to push this one further by using the 5-factor model Fama and French (2015), as well as by carrying out a temporal comparison between 1990-2005 and 2005-2019. The aim is to study the changes that may have occurred in asset management over these periods and to test the effect of this new 5-factor model on the results previously obtained.

Fama and French (2010): *« The challenge is to* ***distinguish skill from luck****. Given the multitude of funds, many have extreme returns by chance. A common approach to this problem is to test for persistence in fund returns, that is, whether past winners continue to produce high returns and losers continue to underperform (see, e.g., Grinblatt and Titman (1992), Carhart (1997)).* ***Persistence tests have an important weakness****. Because they rank funds on short-term past performance, there may be little evidence of persistence because the allocation of funds to winner and loser portfolios is largely based on noise."* Indeed, the most intuitive way to rank fund managers would be to compare them on their performance by period and see if the best remain at the top of the rankings. However, since returns are influenced by white noise, some funds may fortunately show very high returns over a longer or shorter period of time.

Barras et al. (2010) noticed this issue : *«The previous literature, referred to as the standard approach, proposes to measure the number of funds with differential performance by the number of significant funds (Jensen, 1968; W. E. Ferson and Schadt, 1996; W. Ferson and Qian, 2004). Stated differently, it simply counts the number of funds which are located at the tails of the cross-sectional alpha distribution. This approach is problematic, because it does not account for the presence of lucky funds among these significant funds."* and introduced a new risk measure False Discovery Rate (FDR). In this paper we will only have a focus on Fama and French.

Fama and French then proposed an innovative approach, studying the funds over a much longer period but this time not focusing on returns but on alphas, and more specifically on the significance of these alphas. The objective is to calculate the alphas of each fund and then compare them to alphas obtained by simulation. Simulations are used to obtain estimated alpha where there is no longer any abnormal performance due to market uncertainty. Comparison with the alphas obtained, referring to those estimated without simulation on the historical returns of the funds, then makes it possible to see the existence, or not, of the talent of the managers. We will thus estimate the alphas using the three- and five-factor models. In a second step, to determine the simulated alphas, the returns minus their alphas will be regressed over 10,000 simulations. In each simulation the factors belonging to the period studied are randomly selected with discount. Setting alphas to zero on net returns implies a world where all managers have enough talent to produce an expected return that covers their management costs. Thus the simulated results represent, without market fluctuations, the alphas that managers should produce to cover their costs. Comparison with the actual results then allows conclusions to be drawn about the actual performance of mutual funds and the talent of the managers.

# 2 Methodology

## 2.1 Data Engineering

Our research take up the ideas developed by Fama and French (2010). We started from the same database as they did, in order to avoid any bias in the data before we even started the calculations. We therefore chose the database of the mutual funds of the Center for Research in Security Prices.[2] (CRSP). However, our study includes a difference over the observation period, 1990-2019 versus 1984-2006 for Fama and French, in order to compare the evolution of values over time.

We have thus separated our data into two sub-periods 1990-2005, 2005-2019 with the lower bounds included and the upper bounds excluded. Originally our data range extended to 1970, however the data available over this period were too few and too insignificant. As Fama and French pointed out, very few funds reported their monthly results before 1984, which constitutes a strong selection bias. (Elton et al.

[2] http://www.crsp.com/

(2001) discuss CRSP data problems for the period before 1984). We therefore decided, after studying the simulated coefficients and $T_\alpha$, not to take into account the period 1970-1990. Thereafter all funds report monthly data.

We then segregated these funds according to their AuM, while taking up the boundaries of the Fama and French study. We believe that these groups are relevant and make it possible to compare the results. The value of AuM retained for this sorting is the last available value for the fund. Thus, after filtering, the method of which is detailed in the following section, the 6 sub-groups created contain 2735, 652 and 427 funds respectively over the period 1990-2005, then 6252, 1843 and 1157 funds over the second period. Finally, contrary to their study, we did not focus on gross returns, but only on the net returns of mutual funds.

## 2.2 Filtration

Taking all the funds of the CRSP would not be very interesting, as the mix of different types of funds (diversified, index, monetary etc...), or the inclusion of statistical bias such as incubation bias, could make the results obtained corrupted or meaningless. The first bias to be eliminated and which had been identified by Fama and French is the incubation bias. Indeed, during the period of fund creation and during the contribution of the first capital, returns are subject to a non-trivial effect. In order to not suffer from this effect and to be able to compare our results with the initial study, we decided to follow the same method as they did. The returns of a fund are thus taken into account as soon as the amount of AuM has reached at least one time the threshold of 2.5 million. Another problem that Fama and French did not have was the fact that we did not have all the information, such as the types of funds and their investment universe. However, this data is very important here, because three-factor and five-factor models, although adapted to an actively managed equity or diversified fund, have little interest in a money market or index fund that would disrupt our results. Thus, by comparing the curves obtained over the same period with these undesirable funds in our data, it appeared that the difference obtained with those of Fama and French, which served as a benchmark for our study to test the impact of these funds, was substantial and prevented any interpretation of the data. However, since we did not have access to the type of funds, we were able to define a method for selecting them. We tried to characterize their types according to their returns over each period. Thus the funds retained were only selected under conditions.

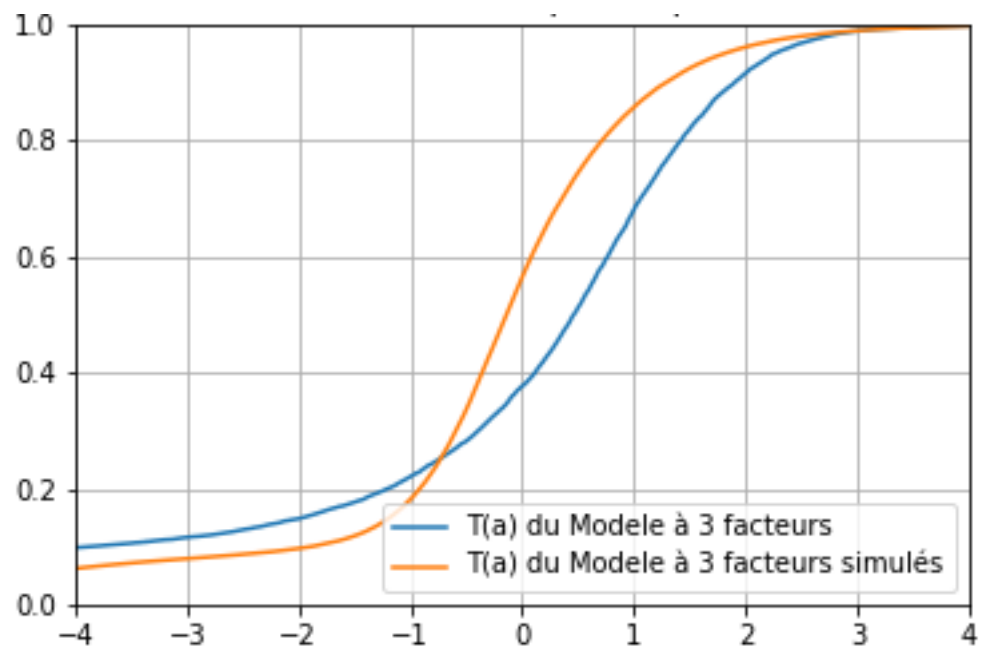

Figure 1: Cumulative density function of $T_\alpha$ 1984-2006 before selection

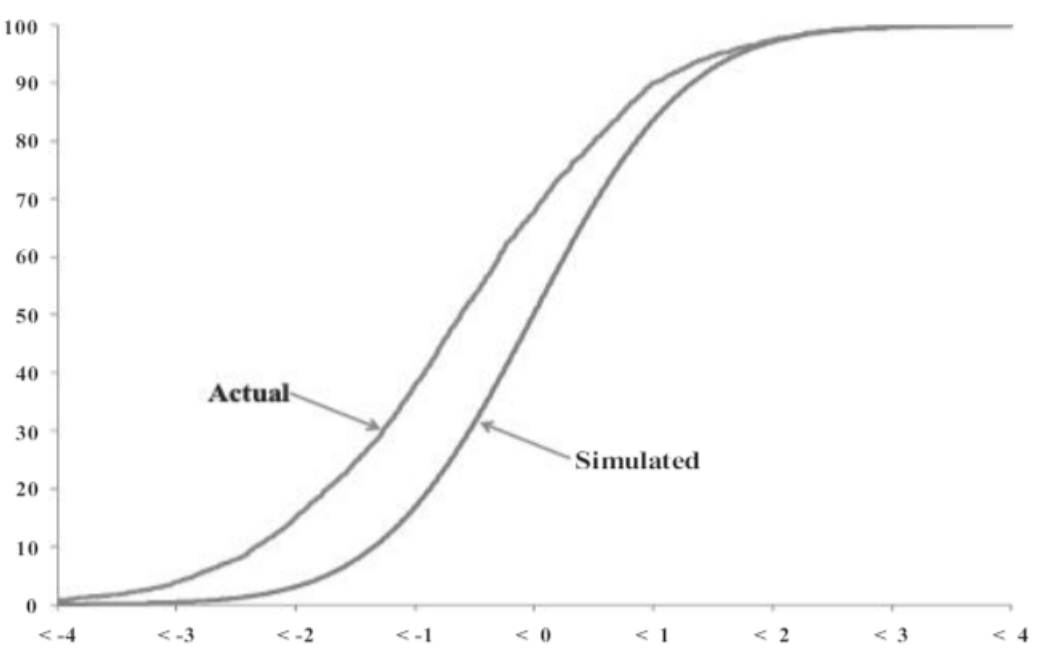

Figure 2: Cumulative density function of $T_\alpha$ 1984-2006 Fama et French

The first selection allow us to eliminate short-term money/bond funds. As these are closely linked to the risk-free rate, in this case 1-month T-Bills, we have regressed the returns of the funds on the risk-free rate. Only funds with a Student's T of the $Rf_t$ coefficient (the risk-free rate) lower than 8 were selected. This value, although important, is in fact easily reached for this type of funds and avoids a too important loss of other funds since the risk-free rate itself is linked to the monetary policy in force. The risk-free rate has an effect on the equity market, which can be appreciated following the measures taken by the central banks. The second selection aims to eliminate index funds as well as highly leveraged funds, although the process was more difficult than for money market funds. Indeed, diversified funds are obviously also strongly correlated to market returns. The solution found to best fit the results of Fama and French is the following condition: $(|\beta|-1 > 0.05)$ or $(|\mathbf{T}_\beta|) < 8)$ and $(|\beta|-1 < 5)$ $|\mathbf{T}_\beta| > 1.95)$.

This condition tests both whether a fund is not an index fund (first part) but also whether it is an equity or diversified fund, and whether it is not a leveraged fund (second part).

Finally, in each period, only funds appearing for at least two years are retained, in order to obtain statistical significance, even if this includes survivor bias .
However, this induces a reduction in the sample and may cause survivorship bias to arise in our series, but we considered this effect less important than that caused by too short series. Thanks to these two conditions, our initial background population has been considerably reduced and the graphical result obtained is as follows :

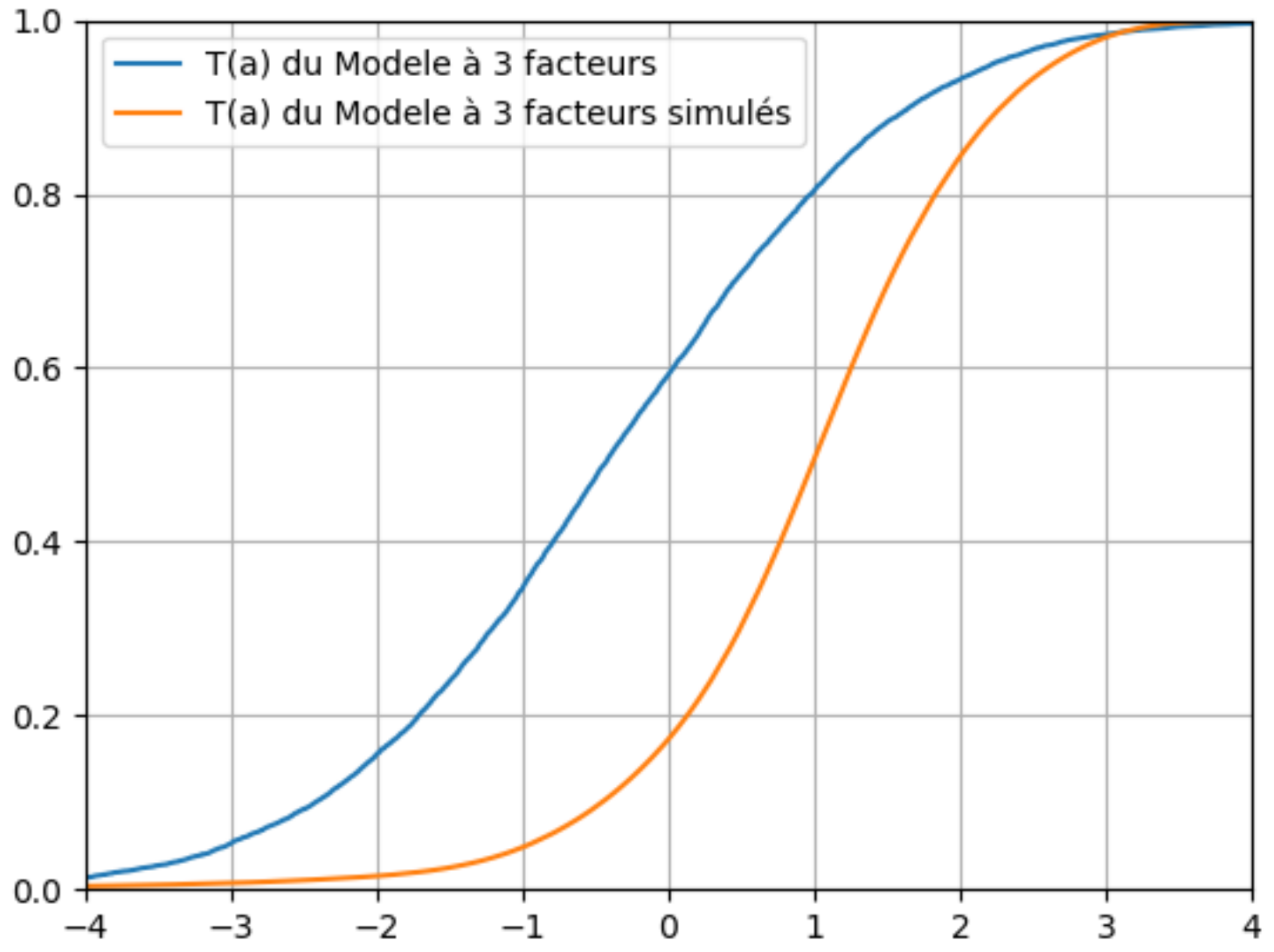

Figure 3: Cumulative density function of T($\alpha$) 1984-2006 all size

In addition to this more obvious similarity, a comparison of the following table with that of Fama and French allows us to see their similarity. This step of filtering the backgrounds would be avoidable with the necessary data, but is here essential to obtain meaningful data for the rest of the study. A study of the $T_\alpha$ values corroborates the fact that our samples are quite similar to those of Fama and French below.

| | 5 Million | | | 250 Million | | | 1 Billion | | |
|---|---|---|---|---|---|---|---|---|---|
| Pct | Sim | Act | %<Act | Sim | Act | %<Act | Sim | Act | %<Act |
| **3-Factor Net Returns** | | | | | | | | | |
| 1 | −2.50 | −3.87 | 0.08 | −2.45 | −3.87 | 0.10 | −2.39 | −4.39 | 0.01 |
| 2 | −2.17 | −3.42 | 0.06 | −2.13 | −3.38 | 0.13 | −2.09 | −3.55 | 0.09 |
| 3 | −1.97 | −3.15 | 0.07 | −1.94 | −3.15 | 0.12 | −1.91 | −3.36 | 0.07 |
| 4 | −1.83 | −2.99 | 0.06 | −1.80 | −3.04 | 0.10 | −1.78 | −3.16 | 0.07 |
| 5 | −1.71 | −2.84 | 0.08 | −1.69 | −2.91 | 0.10 | −1.67 | −2.99 | 0.10 |
| 10 | −1.32 | −2.34 | 0.05 | −1.31 | −2.37 | 0.10 | −1.30 | −2.53 | 0.08 |
| 20 | −0.87 | −1.74 | 0.03 | −0.86 | −1.87 | 0.04 | −0.86 | −1.98 | 0.03 |
| 30 | −0.54 | −1.27 | 0.06 | −0.54 | −1.41 | 0.06 | −0.54 | −1.59 | 0.02 |
| 40 | −0.26 | −0.92 | 0.05 | −0.27 | −1.03 | 0.07 | −0.27 | −1.19 | 0.02 |
| 50 | −0.01 | −0.62 | 0.04 | −0.01 | −0.71 | 0.06 | −0.01 | −0.82 | 0.03 |
| 60 | 0.25 | −0.29 | 0.11 | 0.25 | −0.39 | 0.19 | 0.24 | −0.51 | 0.05 |
| 70 | 0.52 | 0.08 | 0.51 | 0.52 | −0.08 | 0.25 | 0.52 | −0.20 | 0.08 |
| 80 | 0.85 | 0.50 | 3.20 | 0.84 | 0.37 | 1.68 | 0.84 | 0.25 | 0.85 |
| 90 | 1.30 | 1.01 | 8.17 | 1.29 | 0.89 | 5.19 | 1.28 | 0.82 | 4.81 |
| 95 | 1.68 | 1.54 | 30.55 | 1.66 | 1.36 | 14.17 | 1.64 | 1.34 | 17.73 |
| 96 | 1.80 | 1.71 | 40.06 | 1.76 | 1.49 | 17.24 | 1.74 | 1.52 | 26.33 |
| 97 | 1.94 | 1.91 | 49.35 | 1.90 | 1.69 | 25.92 | 1.87 | 1.79 | 42.86 |
| 98 | 2.13 | 2.17 | 58.70 | 2.08 | 1.90 | 30.43 | 2.04 | 2.02 | 50.07 |
| 99 | 2.45 | 2.47 | 57.42 | 2.36 | 2.29 | 43.92 | 2.31 | 2.40 | 63.11 |
| **4-Factor Net Returns** | | | | | | | | | |
| 1 | −2.55 | −3.94 | 0.04 | −2.47 | −3.94 | 0.08 | −2.40 | −4.22 | 0.01 |
| 2 | −2.20 | −3.43 | 0.04 | −2.14 | −3.43 | 0.09 | −2.09 | −3.48 | 0.08 |
| 3 | −2.00 | −3.08 | 0.13 | −1.95 | −3.07 | 0.25 | −1.91 | −3.11 | 0.23 |
| 4 | −1.85 | −2.88 | 0.13 | −1.80 | −2.88 | 0.22 | −1.77 | −2.95 | 0.21 |
| 5 | −1.73 | −2.74 | 0.12 | −1.69 | −2.78 | 0.18 | −1.66 | −2.86 | 0.14 |
| 10 | −1.33 | −2.23 | 0.14 | −1.30 | −2.34 | 0.14 | −1.29 | −2.48 | 0.07 |
| 20 | −0.86 | −1.67 | 0.10 | −0.85 | −1.80 | 0.11 | −0.84 | −1.96 | 0.05 |
| 30 | −0.53 | −1.25 | 0.12 | −0.52 | −1.39 | 0.10 | −0.52 | −1.54 | 0.04 |
| 40 | −0.25 | −0.88 | 0.21 | −0.25 | −1.04 | 0.14 | −0.25 | −1.23 | 0.05 |
| 50 | 0.01 | −0.60 | 0.18 | 0.01 | −0.76 | 0.11 | 0.01 | −0.87 | 0.07 |
| 60 | 0.26 | −0.29 | 0.25 | 0.27 | −0.42 | 0.29 | 0.26 | −0.49 | 0.19 |
| 70 | 0.54 | 0.02 | 0.37 | 0.54 | −0.13 | 0.24 | 0.54 | −0.18 | 0.24 |
| 80 | 0.87 | 0.44 | 1.76 | 0.86 | 0.27 | 0.72 | 0.86 | 0.17 | 0.45 |
| 90 | 1.33 | 1.04 | 10.62 | 1.31 | 0.86 | 4.40 | 1.30 | 0.86 | 7.07 |
| 95 | 1.72 | 1.53 | 23.82 | 1.69 | 1.37 | 14.35 | 1.67 | 1.31 | 14.13 |
| 96 | 1.84 | 1.67 | 28.21 | 1.80 | 1.51 | 18.23 | 1.78 | 1.45 | 17.16 |
| 97 | 1.99 | 1.84 | 31.30 | 1.94 | 1.65 | 18.62 | 1.91 | 1.57 | 17.05 |
| 98 | 2.19 | 2.09 | 39.12 | 2.12 | 1.79 | 15.57 | 2.08 | 1.76 | 18.86 |
| 99 | 2.52 | 2.40 | 36.96 | 2.42 | 2.22 | 29.88 | 2.36 | 2.26 | 42.00 |

Figure 4: Fama et French : Summary table of T($\alpha$) 1984-2006

However, it is obvious that this method does not allow us to have exactly the same sample, which means that some funds polluting our results are wrongly retained, but more importantly, some funds that should be taken into account are not. However, their effects now become more marginal and allow the exploitation of the figures for the parts that will follow. In the end, we only have 2364 funds left between 5 and 250 million for this period, when Fama and French counted 3156.

| | 5 Millions | | | 250 Millions | | | 1 Milliard | | |
|---|---|---|---|---|---|---|---|---|---|
| Pct | Sim | Act | %<Act | Sim | Act | %<Act | Sim | Act | %<Act |
| 3 Facteurs | *2363 Fonds* | | | *583 Fonds* | | | *382 Fonds* | | |
| 1 | -2,65 | -4,50 | 0,26 | -2,98 | -3,69 | 0,72 | -1,39 | -3,16 | 0,23 |
| 2 | -1,84 | -3,68 | 0,43 | -1,60 | -3,18 | 0,91 | -1,12 | -2,76 | 0,25 |
| 3 | -1,46 | -3,44 | 0,52 | -1,23 | -2,91 | 1,03 | -0,95 | -2,30 | 0,27 |
| 4 | -1,23 | -3,28 | 0,58 | -1,03 | -2,65 | 1,15 | -0,79 | -2,16 | 0,28 |
| 5 | -1,05 | -3,13 | 0,66 | -0,89 | -2,51 | 1,22 | -0,63 | -2,05 | 0,29 |
| 10 | -0,51 | -2,53 | 1,11 | -0,43 | -1,84 | 1,66 | -0,03 | -1,48 | 0,78 |
| 20 | 0,06 | -1,83 | 2,03 | 0,17 | -1,20 | 3,11 | 0,54 | -0,87 | 3,49 |
| 30 | 0,42 | -1,37 | 3,36 | 0,54 | -0,73 | 6,50 | 0,86 | -0,36 | 6,87 |
| 40 | 0,70 | -1,01 | 5,31 | 0,83 | -0,23 | 12,79 | 1,10 | 0,13 | 11,98 |
| 50 | 0,94 | -0,63 | 8,65 | 1,08 | 0,18 | 20,33 | 1,33 | 0,49 | 18,62 |
| 60 | 1,18 | -0,24 | 13,96 | 1,33 | 0,56 | 30,54 | 1,56 | 0,86 | 30,28 |
| 70 | 1,44 | 0,21 | 23,59 | 1,60 | 0,95 | 44,86 | 1,85 | 1,15 | 42,27 |
| 80 | 1,74 | 0,73 | 41,34 | 1,93 | 1,40 | 62,88 | 2,20 | 1,55 | 59,41 |
| 90 | 2,18 | 1,34 | 66,21 | 2,39 | 1,96 | 80,73 | 2,67 | 2,19 | 79,71 |
| 95 | 2,53 | 1,90 | 84,06 | 2,73 | 2,40 | 90,13 | 2,99 | 2,66 | 89,89 |
| 96 | 2,62 | 2,02 | 86,93 | 2,81 | 2,54 | 92,44 | 3,07 | 2,90 | 93,77 |
| 97 | 2,74 | 2,21 | 90,56 | 2,91 | 2,65 | 93,90 | 3,15 | 2,95 | 94,51 |
| 98 | 2,89 | 2,37 | 93,13 | 3,02 | 2,84 | 96,30 | 3,25 | 3,01 | 95,25 |
| 99 | 3,12 | 2,66 | 96,35 | 3,19 | 3,18 | 98,95 | 3,40 | 3,39 | 98,96 |
| 5 Facteurs | *2363 Fonds* | | | *583 Fonds* | | | *382 Fonds* | | |
| 1 | -2,50 | -4,53 | 0,26 | -2,80 | -4,22 | 0,56 | -1,42 | -3,86 | 0,15 |
| 2 | -1,76 | -4,05 | 0,34 | -1,64 | -3,75 | 0,72 | -1,08 | -3,40 | 0,20 |
| 3 | -1,43 | -3,77 | 0,41 | -1,34 | -3,46 | 0,81 | -0,87 | -3,20 | 0,22 |
| 4 | -1,23 | -3,58 | 0,46 | -1,17 | -3,12 | 0,90 | -0,71 | -3,16 | 0,22 |
| 5 | -1,07 | -3,34 | 0,54 | -1,03 | -2,94 | 0,95 | -0,58 | -3,00 | 0,23 |
| 10 | -0,60 | -2,64 | 0,90 | -0,56 | -2,40 | 1,17 | -0,12 | -2,22 | 0,29 |
| 20 | -0,07 | -1,88 | 1,78 | 0,03 | -1,49 | 2,41 | 0,42 | -1,34 | 1,18 |
| 30 | 0,30 | -1,41 | 3,08 | 0,42 | -0,78 | 7,39 | 0,76 | -0,86 | 3,04 |
| 40 | 0,59 | -1,00 | 5,54 | 0,72 | -0,23 | 15,00 | 1,05 | -0,18 | 9,28 |
| 50 | 0,86 | -0,57 | 10,50 | 0,98 | 0,32 | 27,16 | 1,31 | 0,36 | 18,48 |
| 60 | 1,12 | -0,13 | 18,44 | 1,25 | 0,72 | 39,93 | 1,60 | 0,72 | 28,48 |
| 70 | 1,40 | 0,35 | 31,60 | 1,55 | 1,12 | 55,18 | 1,93 | 1,21 | 46,24 |
| 80 | 1,73 | 0,83 | 48,99 | 1,92 | 1,56 | 70,31 | 2,34 | 1,60 | 59,99 |
| 90 | 2,21 | 1,47 | 72,42 | 2,51 | 2,14 | 84,32 | 2,90 | 2,29 | 78,92 |
| 95 | 2,63 | 1,97 | 85,70 | 2,96 | 2,49 | 89,79 | 3,28 | 2,93 | 90,56 |
| 96 | 2,75 | 2,11 | 88,29 | 3,08 | 2,64 | 91,58 | 3,38 | 3,03 | 91,94 |
| 97 | 2,89 | 2,25 | 90,61 | 3,21 | 2,78 | 93,21 | 3,49 | 3,13 | 93,25 |
| 98 | 3,08 | 2,47 | 93,48 | 3,37 | 3,12 | 96,31 | 3,64 | 3,30 | 95,20 |
| 99 | 3,36 | 2,79 | 96,32 | 3,60 | 3,64 | 99,13 | 3,84 | 3,57 | 97,57 |

Figure 5: Summary table of T($\alpha$) 1984-2006

## 2.3 Estimation

Here we will discuss factor estimation using the multiple linear regression method. During our study we had to define a function internal to the programming language used to estimate the alphas and betas of each of the Funds. Indeed, the available packages do not allow to perform the large number of regressions in an acceptable time. We therefore had to develop our own tools in order to optimize the regression processing time.

The linear regression model can be expressed such:

$$p(y|x) = \mathcal{N}(y|f(x), \sigma^2) \tag{5}$$

$$\begin{aligned} y &= f(x;\beta) + \epsilon \\ &= x\beta + \epsilon \end{aligned} \tag{6}$$

Where $y \in \mathbb{R}, X \in \mathbb{R}^D$ and $\epsilon \sim \mathcal{N}(0, \sigma^2)$. Our dataset $D := \{(x_1, y_1), \ldots, (x_N, y_N)\}$ with $x_n \in \mathbb{R}^D$ and corresponding output for each observations $\{y_i\}_{i=1}^N = \{y_1, \ldots, y_N\}$.

$$\begin{pmatrix} Y_1 \\ Y_2 \\ \vdots \\ Y_N \end{pmatrix} = \begin{pmatrix} 1\, X_{1,1} & X_{1,2} & X_{1,3} \\ 1\, X_{2,1} & X_{2,2} & X_{2,3} \\ \vdots\ \vdots & \vdots & \vdots \\ 1 X_{N,1} & X_{N,2} & X_{N,3} \end{pmatrix} \begin{pmatrix} \beta_0 \\ \beta_1 \\ \vdots \\ \beta_p \end{pmatrix} + \begin{pmatrix} \varepsilon_1 \\ \varepsilon_2 \\ \vdots \\ \varepsilon_N \end{pmatrix} \tag{7}$$

The objective is to maximise the likelihood to find $\beta^*$.

$$\beta^* = \arg\max_{\beta} p(Y|B, \beta) \tag{8}$$

Where $y \in \mathbb{R}^N$ and $R \in \mathbb{R}^{N \times D}$. The likelihood function associated is given by:

$$p(y|X, \beta) = p(y_1, \ldots, y_n|x_1, \ldots, x_n) = \prod_{i=1}^{N} p(y_i|x_i) = \prod_{i=1}^{N} \mathcal{N}(y_i|x_i\beta), \sigma^2) \tag{9}$$

Maximising the log-likelihood is equivalent to minimize the negative log-likelihood given by:

$$\log p(y|X, \beta) = \log \prod_{i=1}^{N} p(y_i|x_i) = \sum_{i=1}^{N} \log p(y_i|x_i) \tag{10}$$

$$\begin{aligned} L(\beta) &= \sum_{i=1}^{N} (y_i - x_i\beta)^2 \\ &= (y - X\beta)^T (y - X\beta) \\ &= ||Y - X\beta||^2 \end{aligned} \tag{11}$$

$$\begin{aligned} L(\beta) &= \sum_{i=1}^{N} (y_i - x_i\beta)^2 \\ &= (y - X\beta)^T (y - X\beta) \\ &= ||y - X\beta||^2 \end{aligned} \tag{12}$$

$$\begin{aligned} \frac{\partial L(\beta)}{\partial \beta} &\Leftrightarrow X^T X\beta = yX \\ &\Leftrightarrow \beta = (X^T X)^{-1} X y^T \end{aligned} \tag{13}$$

From (12) we obtain a squared-error-loss function to minimize.

$$\beta^* = \frac{1}{2} \underset{\beta}{\text{argmin}}\, ||y - X\beta||^2$$

We have estimated all the coefficients and their $\mathbf{T}_{stat}$. Our results over the period 1990-2005 are avaible below. For the period from 2005 up to 2019 you will be able to find them in the table 2.

Table 1: Coefficients Période 1990-2005

| **Période 1990-2005** | | | | | | | | | | | | | | | | | | |
|---|---|---|---|---|---|---|---|---|---|---|---|---|---|---|---|---|---|---|
| | 5 Million | | | | | | 250 Million | | | | | | 1 Milliard | | | | | |
| Pct | 10 | 25 | 50 | 75 | 90 | Mean | 10 | 25 | 50 | 75 | 90 | Mean | 10 | 25 | 50 | 75 | 90 | Mean |
| Rdmt moyen Funds | | | *2735 Funds* | | | | | | *652 Funds* | | | | | | *427 Funds* | | | |
| Mean rdmt-rf | -0,10% | 0,09% | 0,28% | 0,54% | 0,80% | 0,30% | 0,15% | 0,24% | 0,45% | 0,72% | 0,96% | 0,50% | 0,22% | 0,37% | 0,56% | 0,74% | 0,96% | 0,57% |
| 3 Facteurs | | | *2735 Funds* | | | | | | *652 Funds* | | | | | | *427 Funds* | | | |
| Alpha | -0,006 | -0,003 | -0,001 | 0,001 | 0,002 | -0,001 | -0,003 | -0,002 | 0,000 | 0,001 | 0,003 | 0,000 | -0,002 | -0,001 | 0,000 | 0,002 | 0,004 | 0,001 |
| Rm-rf | 0,083 | 0,235 | 0,707 | 0,968 | 1,114 | 0,630 | 0,089 | 0,321 | 0,804 | 0,987 | 1,129 | 0,696 | 0,106 | 0,588 | 0,841 | 0,983 | 1,084 | 0,734 |
| SMB | -0,147 | -0,067 | 0,024 | 0,230 | 0,540 | 0,118 | -0,140 | -0,044 | 0,049 | 0,305 | 0,571 | 0,143 | -0,164 | -0,052 | 0,039 | 0,206 | 0,458 | 0,101 |
| HML | -0,286 | -0,027 | 0,080 | 0,208 | 0,448 | 0,074 | -0,332 | -0,005 | 0,098 | 0,261 | 0,497 | 0,098 | -0,329 | 0,006 | 0,103 | 0,369 | 0,520 | 0,128 |
| t_Alpha | -2,375 | -1,560 | -0,663 | 0,361 | 1,155 | -0,661 | -1,766 | -0,887 | -0,033 | 0,963 | 1,669 | -0,042 | -1,458 | -0,648 | 0,373 | 1,282 | 1,959 | 0,309 |
| t_Rm-rf | 2,771 | 4,962 | 10,437 | 19,374 | 28,381 | 13,449 | 3,450 | 6,915 | 15,346 | 24,789 | 33,390 | 17,427 | 3,996 | 10,504 | 19,726 | 31,740 | 40,374 | 22,994 |
| t_SMB | -3,407 | -1,454 | 0,566 | 3,178 | 6,315 | 1,067 | -3,511 | -1,159 | 1,057 | 4,167 | 8,709 | 1,684 | -4,914 | -1,670 | 0,869 | 3,799 | 6,398 | 1,084 |
| t_HML | -3,100 | -0,406 | 1,542 | 3,365 | 6,893 | 1,671 | -4,226 | -0,128 | 2,288 | 4,283 | 9,044 | 2,265 | -4,822 | 0,071 | 3,053 | 7,173 | 11,613 | 3,320 |
| 5 Facteurs | | | *2735 Funds* | | | | | | *652 Funds* | | | | | | *427 Funds* | | | |
| Alpha | -0,006 | -0,003 | -0,001 | 0,001 | 0,003 | -0,001 | -0,004 | -0,002 | 0,000 | 0,002 | 0,003 | 0,000 | -0,003 | -0,002 | 0,000 | 0,002 | 0,004 | 0,000 |
| Rm-rf | 0,081 | 0,213 | 0,728 | 0,980 | 1,111 | 0,635 | 0,090 | 0,296 | 0,838 | 1,004 | 1,116 | 0,701 | 0,111 | 0,593 | 0,880 | 1,009 | 1,089 | 0,752 |
| SMB | -0,159 | -0,063 | 0,027 | 0,205 | 0,526 | 0,109 | -0,144 | -0,027 | 0,054 | 0,299 | 0,586 | 0,145 | -0,131 | -0,041 | 0,055 | 0,212 | 0,445 | 0,113 |
| HML | -0,284 | -0,072 | 0,066 | 0,195 | 0,372 | 0,053 | -0,220 | -0,029 | 0,089 | 0,235 | 0,396 | 0,083 | -0,296 | -0,033 | 0,083 | 0,249 | 0,414 | 0,075 |
| RMW | -0,291 | -0,145 | -0,008 | 0,091 | 0,220 | -0,032 | -0,261 | -0,111 | 0,020 | 0,121 | 0,282 | 0,005 | -0,189 | -0,049 | 0,046 | 0,164 | 0,286 | 0,049 |
| CMA | -0,216 | -0,082 | 0,014 | 0,136 | 0,307 | 0,034 | -0,195 | -0,088 | 0,000 | 0,122 | 0,251 | 0,015 | -0,160 | -0,051 | 0,043 | 0,165 | 0,299 | 0,080 |
| t_Alpha | -2,600 | -1,650 | -0,602 | 0,488 | 1,324 | -0,631 | -2,233 | -1,252 | 0,166 | 1,024 | 1,867 | -0,124 | -2,173 | -1,222 | 0,027 | 1,256 | 2,172 | -0,002 |
| t_Rm-rf | 2,144 | 3,987 | 8,713 | 16,692 | 25,260 | 11,603 | 3,042 | 5,529 | 12,799 | 21,950 | 30,105 | 15,342 | 3,278 | 9,208 | 17,723 | 29,512 | 38,366 | 20,674 |
| t_SMB | -3,134 | -1,245 | 0,538 | 2,595 | 5,217 | 0,942 | -3,143 | -0,788 | 0,990 | 3,568 | 7,971 | 1,699 | -3,634 | -0,988 | 1,284 | 3,415 | 6,098 | 1,388 |
| t_HML | -1,729 | -0,636 | 0,895 | 2,211 | 3,832 | 0,921 | -1,855 | -0,389 | 1,426 | 2,814 | 4,615 | 1,327 | -2,681 | -0,374 | 1,590 | 3,572 | 5,829 | 1,642 |
| t_RMW | -2,295 | -1,309 | -0,092 | 1,156 | 2,650 | 0,025 | -2,550 | -1,275 | 0,291 | 1,920 | 3,727 | 0,417 | -2,263 | -0,793 | 0,841 | 2,824 | 4,719 | 1,049 |
| t_CMA | -1,610 | -0,716 | 0,124 | 1,039 | 1,917 | 0,156 | -1,892 | -0,898 | -0,001 | 1,044 | 1,972 | 0,065 | -1,854 | -0,645 | 0,528 | 1,878 | 2,709 | 0,519 |

# 3 Result Analysis

## 3.1 Comparison of 3-factor and 5-factor models over the period 1990-2005

As a reminder, the 1990-2005 period includes 2735, 652 and 427 funds in each size group respectively, making it a statistically significant sample for our study. We therefore first extracted all the coefficients of the two models estimated for each subgroup, which gives us the six sections of the table 1. We have also added a line with the returns (please note that the returns are monthly and not annual).

As we anticipated, the addition of explanatory factors mechanically decreases the significance of the old factors with $\mathbf{T}_x$ the T-distribution of $x$. This is well observed with a decrease in the averages of $\mathbf{T}_{R_m-R_f}$, $\mathbf{T}_{SMB}$ et $\mathbf{T}_{HML}$ respectively we have noted 14%, 12,5% and 45% of decrease. However, this does not take away from the importance of these factors, as the percentile decomposition shows that some of the funds remain significantly sensitive to these factors. The market factor obviously stands out here. As our study mainly focuses on equity funds, their returns are therefore strongly correlated to the market, so even the 10th percentile has a student t-distribution of more than 2.

The two additional factors in the 5-factor model are on average less explanatory, especially for smaller funds. Nevertheless, a look at the percentile also shows that a portion of the funds are well impacted by these factors. Thus, even if the effect is not significant for all funds, this model provides more details for a portion of them representing 20 to 30% of our sample. However, the aim of our study is to look at the funds' alphas.

Having seen the usefulness of these new factors in explaining the returns of certain funds, it is now appropriate to look at the effect on alphas and their Student's T. Mechanically, the effect of the 2 additional factors brings the averages of the alphas and their T-distribution close to zero, but the effect on the values within the sample is more interesting. Indeed, the addition of these factors pushes away the distribution of alphas and their t-distribution between the funds. This means that the interpretation that can be made of the quality of managers or the market randomness effect is altered with a larger effect. For funds between 5 and 250 million in AuM, the 10th and 90th percentile thus diverge by 10 to 20% for alphas and their t-distribution value. This approximation of the averages towards 0 of the alphas is closer to the model of Berk & Green (2004) than the experiment of Fama and French, even if on average the alphas net of costs remain below 0 for all groups. Results **12** also allow to see this gap between the $\mathbf{T}_\alpha$ of the two models for current values.

Funds between 5 and 250 million AuM, the range at 2% (1st to 99th percentile) thus goes from [-3.3 ;2.98] to [-4.06 ; 3.05]. It then becomes interesting to observe the effects on the $T_\alpha$ simulated in our bootstrapping method. It appears that where the 5-factor model leads to a fairly large spread on the lower bound and a more moderate spread on the upper bound, the effect on the simulated values is inverse. Indeed, the bounds spread well but the lower bound moves relatively less, from -1.02 to -1.14, while the upper bound moves relatively more, from 3.07 to 3.55. Moreover, this spread is not only felt at the extreme bounds, but also at all other percentiles of current and simulated $\mathbf{T}_\alpha$ in the sample. The impact of the five-factor model is therefore not negligible for our analysis, and the factors it contains are relevants.

With these movements at the limits, one might think that where Fama and French were talking about discouraging results for investors, those obtained with the five factors would be even more so. However, the third column gives a more global view of the effect. This column, which contains the percentage of simulated $\mathbf{T}_\alpha$ values below the percentile of the current value, shows a surprising result for funds with AuM of between 5 million and 250 million (note that this column, which has the same name, does not represent the same thing as the one used by Fama and French in their study). These funds benefit from the 5-factor model, so that a larger part of the simulated $\mathbf{T}_\alpha$ becomes lower than the real value of each percentile compared to the 3-factor model. However, this effect is less for intermediate funds (250 million to 1 billion) where it only benefits funds belonging to the 40th percentile and above. For large funds, the effect is the complete opposite and therefore negative on the analysis of their $\mathbf{T}_\alpha$. Thus, it does not benefit the analysis made on the fund manager.

Finally, even if these effects on this third column are quite important, they do not change the results obtained by Fama and French. Indeed, the $\mathbf{T}_\alpha$. are always lower than the simulated values. Moreover, the small funds, which benefit from the model, actually performed much worse than the large funds in the test performed with the three-factor model. The gap between the funds by size is therefore narrowing, but does not reverse the trend. We can speak here of a correction rather than a real change. Finally, it should be noted that funds over 1 billion 3% were able to produce a net benchmark adjusted return that covers and exceeds their costs with the three-factor model, whereas they are now only one percent for the five-factor model.

| Période 1990-2005 | | | | | | | | | |
|---|---|---|---|---|---|---|---|---|---|
| | 5 Millions | | | 250 Millions | | | 1 Milliard | | |
| Pct | Sim | Act | %<Act | Sim | Act | %<Act | Sim | Act | %<Act |
| 3 Facteurs | *2735 Fonds* | | | *652 Fonds* | | | *427 Fonds* | | |
| 1 | -1,33 | -4,22 | 0,18 | -1,02 | -3,30 | 0,15 | -1,03 | -2,80 | 0,00 |
| 2 | -0,95 | -3,67 | 0,19 | -0,67 | -3,15 | 0,15 | -0,76 | -2,20 | 0,01 |
| 3 | -0,73 | -3,39 | 0,19 | -0,44 | -3,01 | 0,16 | -0,49 | -2,06 | 0,02 |
| 4 | -0,58 | -3,19 | 0,19 | -0,27 | -2,59 | 0,16 | -0,26 | -1,95 | 0,03 |
| 5 | -0,45 | -3,03 | 0,20 | -0,14 | -2,41 | 0,17 | -0,09 | -1,88 | 0,04 |
| 10 | -0,03 | -2,37 | 0,26 | 0,24 | -1,77 | 0,28 | 0,36 | -1,46 | 0,24 |
| 20 | 0,43 | -1,80 | 0,47 | 0,62 | -1,14 | 0,78 | 0,76 | -0,87 | 1,55 |
| 30 | 0,73 | -1,37 | 0,93 | 0,87 | -0,71 | 1,85 | 1,04 | -0,49 | 3,01 |
| 40 | 0,98 | -1,01 | 1,78 | 1,09 | -0,38 | 3,36 | 1,28 | -0,11 | 4,87 |
| 50 | 1,21 | -0,66 | 3,42 | 1,31 | -0,03 | 6,02 | 1,50 | 0,37 | 10,28 |
| 60 | 1,44 | -0,27 | 6,76 | 1,53 | 0,44 | 14,44 | 1,71 | 0,75 | 19,62 |
| 70 | 1,68 | 0,11 | 12,44 | 1,76 | 0,80 | 26,94 | 1,94 | 1,07 | 31,32 |
| 80 | 1,96 | 0,60 | 25,27 | 2,03 | 1,16 | 42,88 | 2,19 | 1,50 | 50,08 |
| 90 | 2,35 | 1,16 | 47,67 | 2,38 | 1,67 | 66,30 | 2,48 | 1,96 | 70,93 |
| 95 | 2,66 | 1,69 | 70,26 | 2,63 | 2,07 | 81,51 | 2,68 | 2,56 | 92,19 |
| 96 | 2,75 | 1,85 | 76,47 | 2,70 | 2,22 | 85,89 | 2,73 | 2,72 | 95,79 |
| 97 | 2,86 | 2,06 | 83,13 | 2,78 | 2,36 | 89,53 | 2,79 | 2,95 | 98,63 |
| 98 | 3,00 | 2,26 | 88,05 | 2,89 | 2,58 | 94,17 | 2,88 | 3,05 | 99,19 |
| 99 | 3,23 | 2,56 | 93,72 | 3,07 | 2,98 | 98,58 | 3,01 | 3,30 | 99,81 |
| 5 Facteurs | *2735 Fonds* | | | *652 Fonds* | | | *427 Fonds* | | |
| 1 | -1,37 | -4,46 | 0,18 | -1,14 | -4,06 | 0,15 | -0,96 | -3,84 | 0,00 |
| 2 | -1,01 | -4,05 | 0,18 | -0,80 | -3,72 | 0,15 | -0,65 | -3,39 | 0,00 |
| 3 | -0,80 | -3,73 | 0,19 | -0,60 | -3,42 | 0,15 | -0,46 | -3,14 | 0,00 |
| 4 | -0,66 | -3,44 | 0,19 | -0,45 | -3,10 | 0,16 | -0,31 | -3,03 | 0,00 |
| 5 | -0,55 | -3,22 | 0,20 | -0,34 | -2,88 | 0,16 | -0,19 | -2,79 | 0,00 |
| 10 | -0,17 | -2,60 | 0,24 | 0,04 | -2,23 | 0,20 | 0,21 | -2,17 | 0,03 |
| 20 | 0,28 | -1,86 | 0,46 | 0,46 | -1,53 | 0,47 | 0,65 | -1,53 | 0,26 |
| 30 | 0,59 | -1,40 | 0,94 | 0,75 | -0,97 | 1,41 | 0,96 | -0,94 | 1,05 |
| 40 | 0,86 | -0,99 | 2,08 | 1,00 | -0,52 | 3,47 | 1,23 | -0,52 | 2,62 |
| 50 | 1,11 | -0,60 | 4,49 | 1,23 | 0,17 | 12,44 | 1,49 | 0,03 | 7,37 |
| 60 | 1,36 | -0,18 | 9,75 | 1,48 | 0,49 | 20,78 | 1,77 | 0,57 | 17,85 |
| 70 | 1,63 | 0,26 | 19,47 | 1,76 | 0,87 | 34,62 | 2,06 | 1,03 | 32,64 |
| 80 | 1,95 | 0,70 | 34,17 | 2,10 | 1,20 | 48,42 | 2,40 | 1,49 | 49,76 |
| 90 | 2,39 | 1,32 | 58,48 | 2,58 | 1,87 | 73,42 | 2,81 | 2,17 | 73,38 |
| 95 | 2,76 | 1,82 | 76,08 | 2,95 | 2,32 | 85,15 | 3,11 | 2,60 | 85,18 |
| 96 | 2,86 | 1,95 | 79,98 | 3,04 | 2,46 | 87,95 | 3,19 | 2,72 | 87,96 |
| 97 | 2,99 | 2,10 | 83,79 | 3,16 | 2,69 | 91,77 | 3,29 | 2,78 | 89,39 |
| 98 | 3,15 | 2,23 | 86,78 | 3,32 | 2,83 | 93,64 | 3,41 | 3,12 | 95,15 |
| 99 | 3,40 | 2,74 | 94,78 | 3,55 | 3,05 | 96,09 | 3,59 | 3,73 | 99,43 |

Figure 6: Results over the period 1990-2005

Graphically, this can be observed by the tightening of the curves on the graph for small funds 12, whereas for funds managing more than 1 billion AuM, the opposite effect is observed with a curve of realised $\mathbf{T}_\alpha$ that no longer, or almost no longer, falls below that of simulated $\mathbf{T}_\alpha$.

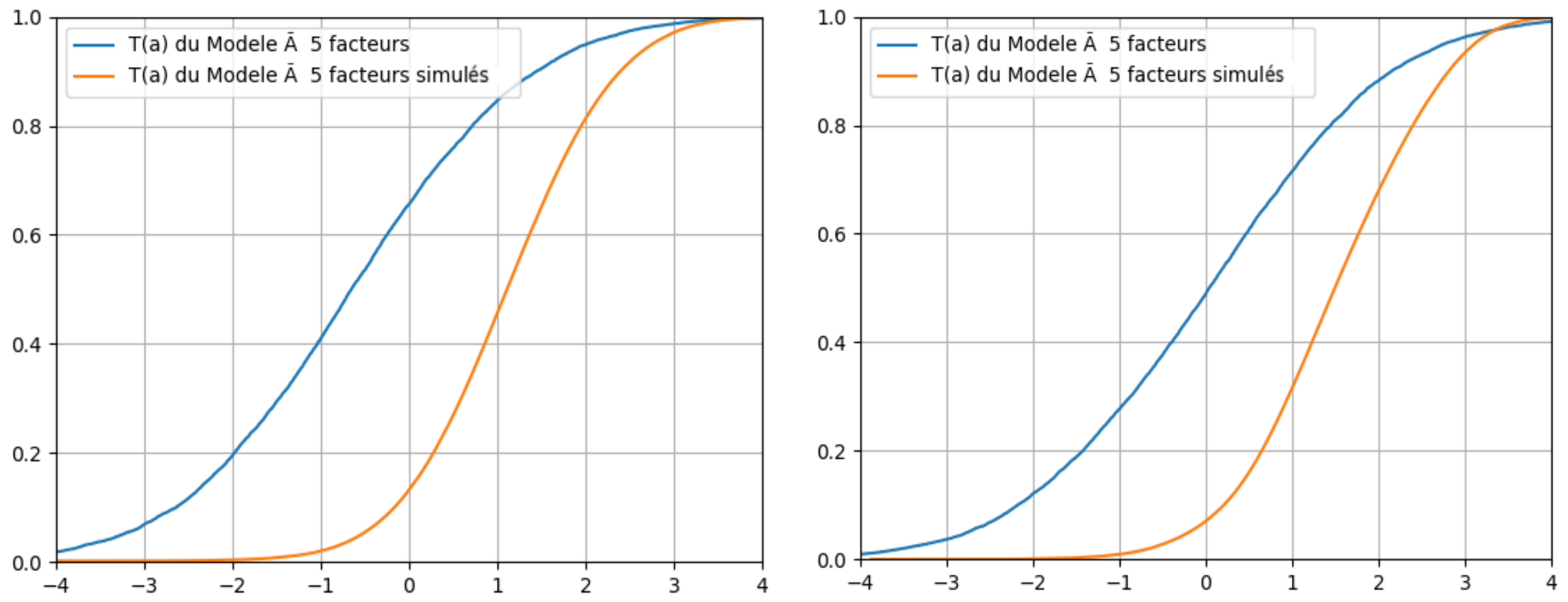

(a) Result for 5 Million AuM Funds (1990-2005)

(b) Result for 1 Billion AuM Funds (1990-2005)

Figure 7: Comparison of the cumulative distribution functions

## 3.2 Analysis over the periods 1990-2005 and 2005-2019

For the rest of the study and having seen the interest in using the five-factor model previously, we will now only use this one. To find the complete tables with the three-factor model, please refer to the appendices. The new period we are now introducing, 2005-2019, has some differences with the previous section. Firstly, the number of funds surveyed has increased by more than 140%, with a greater increase for funds with more than 250 million AuM and those in the over 1 billion category (180% and 170% respectively in each category).

The difference stems more from the management style that has developed since those years. Indeed, although Fama and French developed the 3-factor model in 1993, and other models such as Carhart (1997) followed, their application their application in the industry started around 2004. The interest of this part will be to observe the effects of these new management methods on the results obtained during this period. Since funds are now judged against these models and no longer against the CAPM (Capital Asset Pricing Model), their investment strategies should have had an impact on the regression results. Table 2 of coefficients allow us to see differences. Although fund returns are broadly similar to the previous period (except for an improvement of 9 basis points for funds with less than 250 million AuM, which is not sufficient to catch up with the averages of the other two groups), the other coefficients show more differences. Thus, with the exception of the market and investment coefficient, which remains constant with a rising Student T, the SMB, HML and RMW factors see their average fall drastically, as well as their significance. However, the 20% range still shows a strong contrast between the funds, even if it is less than in the previous period.

Table 2: Coefficients over the period 2005-2019

| | | | | | | | | | | | | | | | | | | |
|---|---|---|---|---|---|---|---|---|---|---|---|---|---|---|---|---|---|---|
| | **Période 2005-2019** | | | | | | | | | | | | | | | | | |
| | 5 Millions | | | | | | 250 Millions | | | | | | 1 Billion | | | | | |
| Pct | 10 | 25 | 50 | 75 | 90 | Mean | 10 | 25 | 50 | 75 | 90 | Mean | 10 | 25 | 50 | 75 | 90 | Mean |
| Rdmt moyen Funds | *6252 Funds* | | | | | | *1843 Funds* | | | | | | *1157 Funds* | | | | | |
| Mean rdmt-rf | -0,05% | 0,18% | 0,40% | 0,65% | 0,89% | 0,39% | 0,13% | 0,29% | 0,49% | 0,71% | 0,92% | 0,49% | 0,21% | 0,35% | 0,53% | 0,73% | 0,93% | 0,56% |
| 3 Facteurs | *6252 Funds* | | | | | | *1843 Funds* | | | | | | *1157 Funds* | | | | | |
| Alpha | -0,005 | -0,003 | -0,001 | 0,001 | 0,002 | -0,001 | -0,004 | -0,002 | 0,000 | 0,001 | 0,002 | -0,001 | -0,003 | -0,001 | 0,000 | 0,001 | 0,002 | 0,000 |
| Rm-rf | 0,104 | 0,367 | 0,811 | 1,025 | 1,156 | 0,701 | 0,097 | 0,393 | 0,810 | 1,006 | 1,146 | 0,709 | 0,100 | 0,408 | 0,781 | 0,978 | 1,133 | 0,699 |
| SMB | -0,221 | -0,108 | -0,028 | 0,131 | 0,550 | 0,057 | -0,214 | -0,112 | -0,032 | 0,086 | 0,509 | 0,043 | -0,205 | -0,123 | -0,046 | 0,014 | 0,265 | -0,005 |
| HML | -0,324 | -0,129 | -0,027 | 0,072 | 0,229 | -0,032 | -0,296 | -0,110 | -0,026 | 0,064 | 0,201 | -0,033 | -0,219 | -0,087 | -0,022 | 0,060 | 0,172 | -0,023 |
| t_Alpha | -2,415 | -1,542 | -0,480 | 0,551 | 1,501 | -0,470 | -2,092 | -1,155 | -0,089 | 0,980 | 1,886 | -0,082 | -1,819 | -0,922 | 0,159 | 1,294 | 2,194 | 0,195 |
| t_Rm-rf | 2,875 | 5,645 | 12,466 | 25,447 | 36,547 | 16,651 | 3,463 | 6,741 | 15,481 | 29,613 | 40,866 | 19,556 | 3,768 | 7,447 | 19,423 | 34,187 | 50,883 | 23,536 |
| t_SMB | -2,473 | -1,517 | -0,520 | 1,235 | 6,573 | 0,906 | -2,866 | -1,686 | -0,669 | 1,037 | 7,789 | 0,975 | -3,636 | -2,006 | -0,995 | 0,268 | 4,349 | 0,157 |
| t_HML | -3,827 | -1,865 | -0,491 | 1,059 | 2,909 | -0,437 | -4,342 | -1,876 | -0,538 | 1,083 | 3,403 | -0,408 | -3,653 | -1,747 | -0,503 | 1,430 | 3,950 | -0,143 |
| 5 Facteurs | *6252 Funds* | | | | | | *1843 Funds* | | | | | | *1157 Funds* | | | | | |
| Alpha | -0,005 | -0,002 | -0,001 | 0,001 | 0,003 | -0,001 | -0,004 | -0,001 | 0,000 | 0,001 | 0,002 | 0,000 | -0,003 | -0,001 | 0,000 | 0,001 | 0,002 | 0,000 |
| Rm-rf | 0,099 | 0,362 | 0,805 | 1,015 | 1,124 | 0,691 | 0,095 | 0,388 | 0,813 | 0,998 | 1,108 | 0,700 | 0,100 | 0,402 | 0,770 | 0,975 | 1,090 | 0,692 |
| SMB | -0,220 | -0,106 | -0,031 | 0,118 | 0,560 | 0,055 | -0,221 | -0,113 | -0,035 | 0,085 | 0,525 | 0,039 | -0,206 | -0,122 | -0,047 | 0,016 | 0,261 | -0,006 |
| HML | -0,273 | -0,105 | -0,009 | 0,103 | 0,272 | 0,005 | -0,243 | -0,097 | -0,005 | 0,090 | 0,223 | -0,004 | -0,200 | -0,069 | -0,003 | 0,088 | 0,170 | 0,000 |
| RMW | -0,267 | -0,107 | -0,007 | 0,089 | 0,212 | -0,022 | -0,256 | -0,101 | -0,007 | 0,073 | 0,176 | -0,028 | -0,193 | -0,074 | -0,004 | 0,079 | 0,160 | -0,012 |
| CMA | -0,485 | -0,248 | -0,072 | 0,035 | 0,173 | -0,124 | -0,430 | -0,214 | -0,061 | 0,037 | 0,173 | -0,094 | -0,389 | -0,173 | -0,048 | 0,039 | 0,150 | -0,077 |
| t_Alpha | -2,252 | -1,427 | -0,397 | 0,651 | 1,563 | -0,371 | -1,978 | -1,037 | -0,030 | 1,061 | 1,924 | 0,006 | -1,706 | -0,846 | 0,168 | 1,293 | 2,261 | 0,248 |
| t_Rm-rf | 2,619 | 5,310 | 11,444 | 23,573 | 34,143 | 15,453 | 3,233 | 6,419 | 14,440 | 27,751 | 38,616 | 18,353 | 3,417 | 7,077 | 18,364 | 32,299 | 47,767 | 22,294 |
| t_SMB | -2,334 | -1,441 | -0,491 | 1,196 | 6,487 | 0,922 | -2,759 | -1,670 | -0,676 | 1,000 | 7,521 | 0,986 | -3,246 | -1,974 | -1,003 | 0,287 | 4,039 | 0,191 |
| t_HML | -2,744 | -1,349 | -0,130 | 1,217 | 2,866 | -0,027 | -2,933 | -1,482 | -0,090 | 1,212 | 3,119 | -0,036 | -2,524 | -1,335 | -0,058 | 1,412 | 3,321 | 0,150 |
| t_RMW | -1,672 | -0,909 | -0,075 | 0,782 | 1,513 | -0,073 | -1,871 | -0,954 | -0,076 | 0,810 | 1,635 | -0,099 | -1,718 | -0,822 | -0,064 | 0,884 | 1,896 | 0,015 |
| t_CMA | -2,655 | -1,669 | -0,650 | 0,305 | 1,130 | -0,690 | -2,732 | -1,676 | -0,650 | 0,412 | 1,407 | -0,629 | -2,772 | -1,644 | -0,593 | 0,439 | 1,598 | -0,516 |

The alphas of the funds subject to our analysis moved slightly. We also noticed intervals between funds decreased while their $\mathbf{T}_\alpha$ increased. This may be a sign of these new management styles, but the focus on results between simulated $\mathbf{T}_\alpha$ and real $\mathbf{T}_\alpha$ shows this more clearly in the table 13. Graphically, the current $\mathbf{T}_\alpha$ curve shift to the right of the previous period while the simulated $\mathbf{T}_\alpha$ curve moves to the left. In the end, this observed gap between the current and simulated $\mathbf{T}_\alpha$ curves is still present but has narrowed between periods.

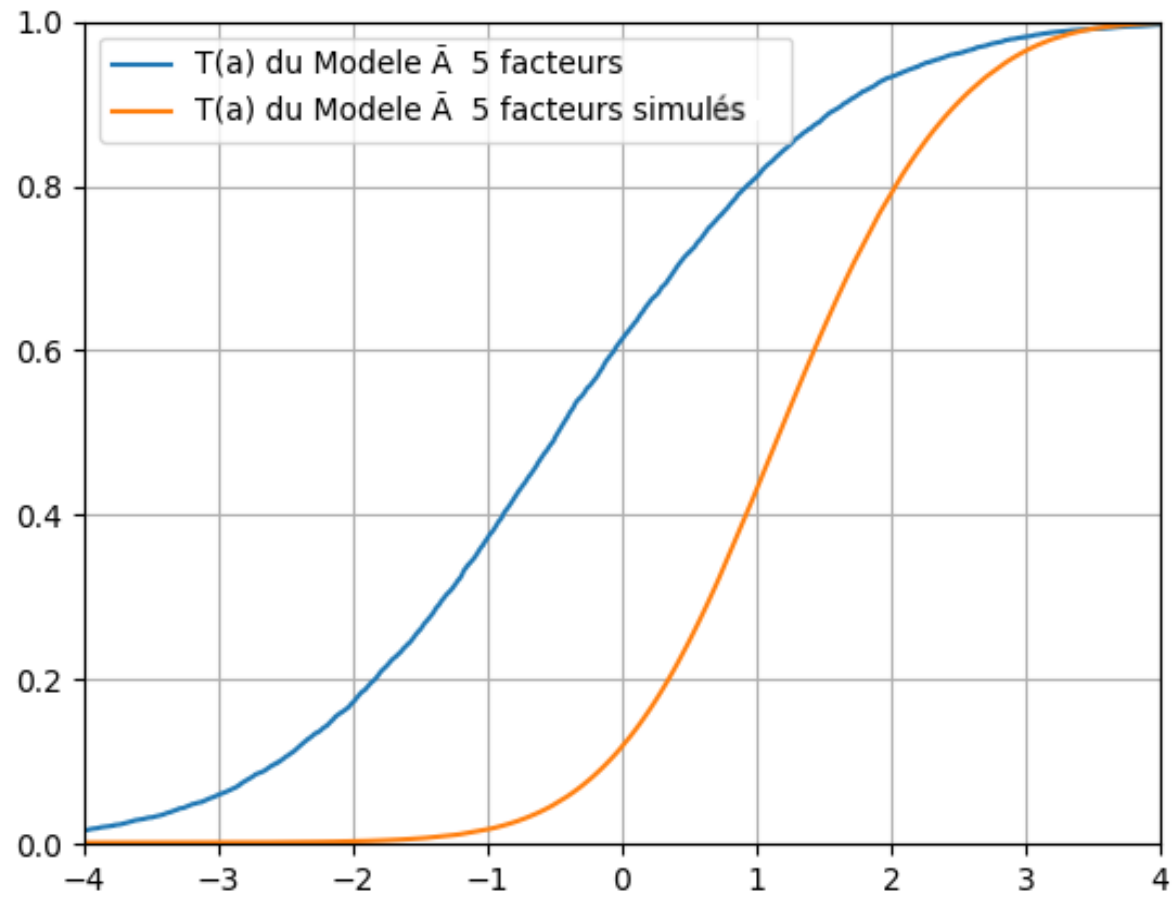

(a) All group of funds (1990-2005)

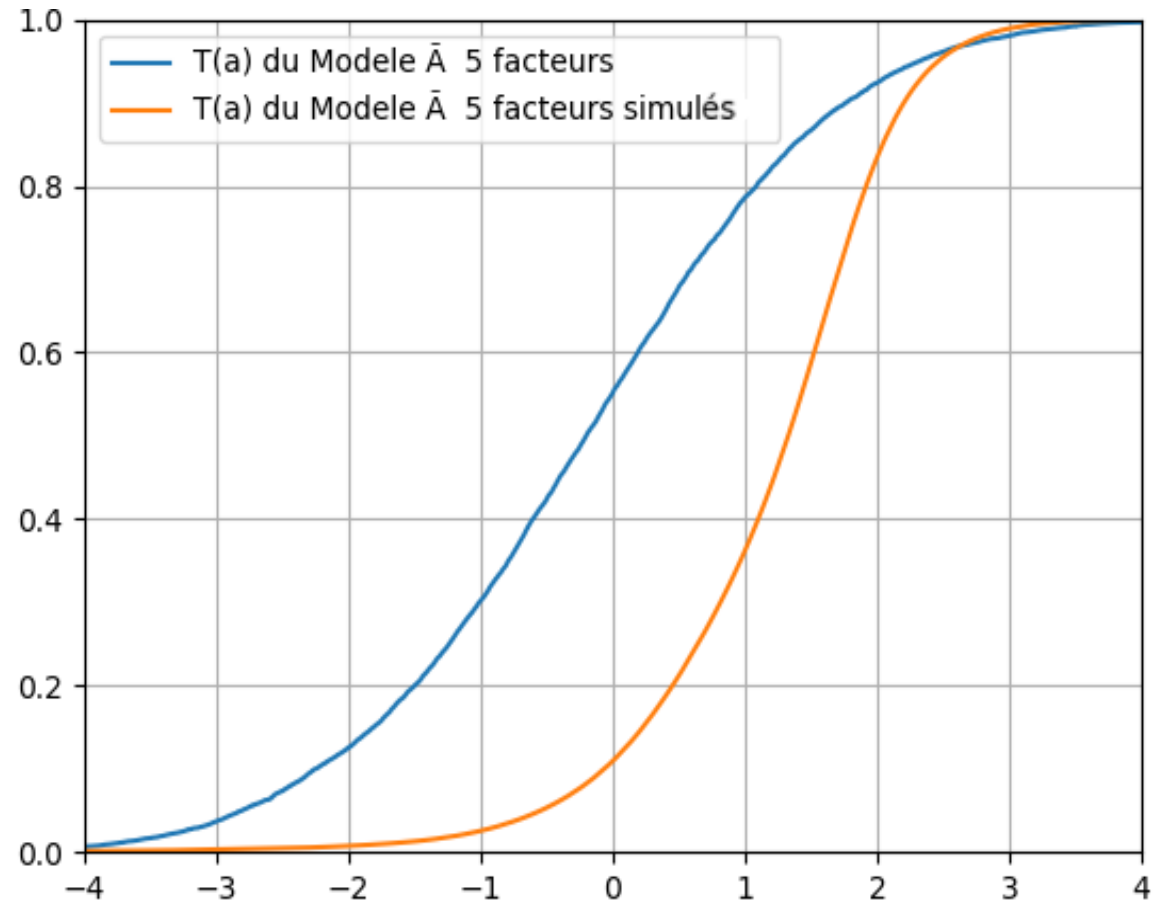

(b) All group of funds (2005-2019)

Figure 8: Comparison of the cumulative distribution functions over the two periods

The observations made earlier are confirmed, as $\mathbf{T}_\alpha$ are higher for funds between 250 million and 1 billion under management. The interval at 2% going from [-4.06 ; 3.5] to [-3.3 ; 3.45], even though the simulated $\mathbf{T}_\alpha$ have seen their interval go from [-1.14 ; 3.55] to [-1.67 ; 3.01]. The column containing for each percentile the percentage of simulated $\mathbf{T}_\alpha$ lower than real $\mathbf{T}_\alpha$ confirms this, with increasing values at all levels. These results are logically in line with the new strategies of funds that are now seeking to maximise their alpha in this new context and no longer with the CAPM.
However, observations made by Fama and French remain partially true, as the curves have retained their relative positions, with the vast majority of managers still doing worse than the simulated values, and not producing a return that covers their cost. However, some of them stand out: 2% of fund managers with between 5 and 250 million AuM, and 5% of those with more than 250 million or more than one billion exceed the simulated values. This means that these managers have been able to produce a net adjusted benchmark return that manages to cover their costs and even generate a surplus.

This is a major difference with the study from the previous period, as well as with the study conducted by Fama French, which sheds light on how these new models are taken into consideration in asset management industry. This gap is also visible on graphics 30 and 31. It can be seen that for intermediate and large funds, the curve of realised $\mathbf{T}_\alpha$ falls below that of simulated $\mathbf{T}_\alpha$ on the right-hand side of the graph.

## 3.3 Comparison of funds by AuM over the period 2005-2019

In this section data used comes from table 2 & 13 please refer to the appendices. For an investor looking for a fund, the decision is based on the analysis of many factors such as the track record, the management type, the type of assets, the investment horizon, the reliability of the audit and accounting bodies, the BCP (Business Continuity Plan), the investment strategy but also the size of the fund. The aim here is to study the effect of fund size on returns, using the five-factor model and our simulations.

By first looking at the estimation of the coefficients for each group of funds, a difference first appears on the SMB. This means that large funds with a smaller SMB favour Large Caps over Small Caps. The market coefficient is similar between funds of different sizes, with funds generally having a similar target Beta. The only notable difference is in the $\mathbf{T}_{\mathrm{R_m - R_f}}$ higher with fund size. This is due to the greater portfolio diversification achieved by the larger funds, which is a strong point to consider from a risk management perspective for investors.

| Période2005-2020 | | | | | | | | | |
|---|---|---|---|---|---|---|---|---|---|
| | 5 Millions | | | 250 Millions | | | 1 Milliard | | |
| Pct | Sim | Act | %<Act | Sim | Act | %<Act | Sim | Act | %<Act |
| 3 Facteurs | | *6252 Fonds* | | | *1843 Fonds* | | | *1157 Fonds* | |
| 1 | -1,71 | -3,85 | 0,16 | -1,69 | -3,46 | 0,14 | -1,84 | -3,19 | 0,43 |
| 2 | -1,18 | -3,42 | 0,19 | -1,11 | -3,16 | 0,24 | -1,17 | -2,84 | 0,53 |
| 3 | -0,92 | -3,19 | 0,23 | -0,78 | -2,91 | 0,35 | -0,80 | -2,57 | 0,62 |
| 4 | -0,72 | -3,01 | 0,27 | -0,52 | -2,70 | 0,45 | -0,45 | -2,43 | 0,68 |
| 5 | -0,57 | -2,86 | 0,30 | -0,30 | -2,60 | 0,49 | -0,18 | -2,29 | 0,74 |
| 10 | -0,01 | -2,42 | 0,44 | 0,33 | -2,09 | 0,69 | 0,42 | -1,82 | 1,01 |
| 20 | 0,51 | -1,79 | 0,92 | 0,80 | -1,42 | 1,38 | 0,92 | -1,19 | 1,94 |
| 30 | 0,84 | -1,31 | 1,62 | 1,16 | -0,90 | 2,61 | 1,31 | -0,70 | 3,27 |
| 40 | 1,15 | -0,89 | 3,13 | 1,41 | -0,46 | 4,26 | 1,55 | -0,26 | 4,67 |
| 50 | 1,40 | -0,48 | 5,66 | 1,59 | -0,09 | 6,14 | 1,71 | 0,16 | 7,03 |
| 60 | 1,59 | -0,10 | 8,95 | 1,73 | 0,31 | 9,73 | 1,84 | 0,60 | 13,25 |
| 70 | 1,77 | 0,30 | 14,99 | 1,87 | 0,73 | 18,20 | 1,96 | 1,06 | 23,09 |
| 80 | 1,94 | 0,81 | 29,07 | 2,01 | 1,22 | 32,05 | 2,09 | 1,54 | 39,44 |
| 90 | 2,18 | 1,50 | 55,06 | 2,22 | 1,89 | 71,32 | 2,31 | 2,19 | 85,72 |
| 95 | 2,44 | 2,14 | 88,79 | 2,46 | 2,55 | 95,99 | 2,59 | 2,74 | 96,36 |
| 96 | 2,53 | 2,32 | 93,27 | 2,55 | 2,69 | 97,14 | 2,69 | 2,85 | 97,14 |
| 97 | 2,65 | 2,52 | 95,86 | 2,67 | 2,83 | 97,99 | 2,83 | 3,08 | 98,51 |
| 98 | 2,80 | 2,79 | 97,96 | 2,83 | 2,98 | 98,78 | 2,99 | 3,53 | 99,74 |
| 99 | 3,01 | 3,30 | 99,67 | 3,04 | 3,53 | 99,87 | 3,19 | 4,52 | 99,91 |
| 5 Facteurs | | *6252 Fonds* | | | *1843 Fonds* | | | *1157 Fonds* | |
| 1 | -1,72 | -3,62 | 0,17 | -1,67 | -3,30 | 0,19 | -1,84 | -2,91 | 0,52 |
| 2 | -1,21 | -3,25 | 0,21 | -1,05 | -3,02 | 0,29 | -1,10 | -2,71 | 0,59 |
| 3 | -0,94 | -3,01 | 0,25 | -0,73 | -2,76 | 0,40 | -0,72 | -2,47 | 0,68 |
| 4 | -0,76 | -2,85 | 0,28 | -0,51 | -2,59 | 0,47 | -0,45 | -2,30 | 0,75 |
| 5 | -0,62 | -2,71 | 0,32 | -0,34 | -2,48 | 0,51 | -0,26 | -2,20 | 0,79 |
| 10 | -0,16 | -2,25 | 0,53 | 0,14 | -1,98 | 0,75 | 0,24 | -1,71 | 1,11 |
| 20 | 0,35 | -1,66 | 1,08 | 0,64 | -1,29 | 1,50 | 0,76 | -1,06 | 2,11 |
| 30 | 0,71 | -1,20 | 2,01 | 0,98 | -0,78 | 2,83 | 1,12 | -0,62 | 3,33 |
| 40 | 1,01 | -0,77 | 3,94 | 1,23 | -0,39 | 4,68 | 1,36 | -0,22 | 5,25 |
| 50 | 1,25 | -0,40 | 7,03 | 1,43 | -0,03 | 7,82 | 1,55 | 0,17 | 9,03 |
| 60 | 1,45 | -0,02 | 12,21 | 1,60 | 0,40 | 14,56 | 1,72 | 0,64 | 17,31 |
| 70 | 1,65 | 0,42 | 21,55 | 1,77 | 0,82 | 24,86 | 1,88 | 1,07 | 28,36 |
| 80 | 1,87 | 0,89 | 35,86 | 1,96 | 1,29 | 42,71 | 2,06 | 1,50 | 47,27 |
| 90 | 2,16 | 1,56 | 65,52 | 2,23 | 1,92 | 78,40 | 2,33 | 2,26 | 87,86 |
| 95 | 2,43 | 2,12 | 88,79 | 2,47 | 2,49 | 95,19 | 2,60 | 2,74 | 96,50 |
| 96 | 2,51 | 2,27 | 92,39 | 2,55 | 2,62 | 96,75 | 2,68 | 2,90 | 97,69 |
| 97 | 2,62 | 2,47 | 95,57 | 2,65 | 2,81 | 98,09 | 2,80 | 3,20 | 98,99 |
| 98 | 2,76 | 2,77 | 98,04 | 2,79 | 3,06 | 99,15 | 2,95 | 3,60 | 99,71 |
| 99 | 3,00 | 3,25 | 99,55 | 3,01 | 3,45 | 99,80 | 3,20 | 4,43 | 99,91 |

Figure 9: Résultats période 2005-2019

The investment factor (CMA) is also higher and more significant for large funds, so even though they invest primarily in Large Caps, they still favour companies that have a more aggressive investment strategy. These companies are generally higher growth companies, known as Growths, and over the last ten years these companies have shown a 50% growth differential with Value stocks, partly due to the low interest rate environment. These differences in factors thus lead to a difference in the net returns of the funds, with higher performance for the larger funds and a less volatile distribution between them.

Finally, it appears in the table of coefficients that the alphas of the funds are also increasing with their size, except that the 90th percentile of the smallest funds is slightly higher than that of the other funds. The alphas are also more dispersed over the smaller funds, but, above all, their Student T's are lower when they are positive and, conversely, higher in absolute value when they are negative.

However, it is the study of simulated values that will once again shed light on this trend. Indeed, it appears that funds between 5 and 250 million show a much larger gap between their real and simulated $T_\alpha$ than their peers with more AuM. It therefore seems that these funds have not implemented the strategies used by larger funds, and make less use of multifactor models. However, some of these funds still outperformed, 2% (versus 5% for other sizes), by managing to produce a positive benchmark adjusted return.

Finally, the difference in the column of the percentages of simulated $T_\alpha$ lower than actual $T_\alpha$ also shows a much less flagrant decrease than for the other groups of funds between the periods 1990-2005 and 2005-2019.

For an investor, the choice of larger funds, even if it is not a guarantee of higher net returns, therefore seems in hope to be a wiser choice. However, the minimum units to enter certain large funds do not allow all investors to enter them, and explain the existence of smaller funds even if they present a higher risk.

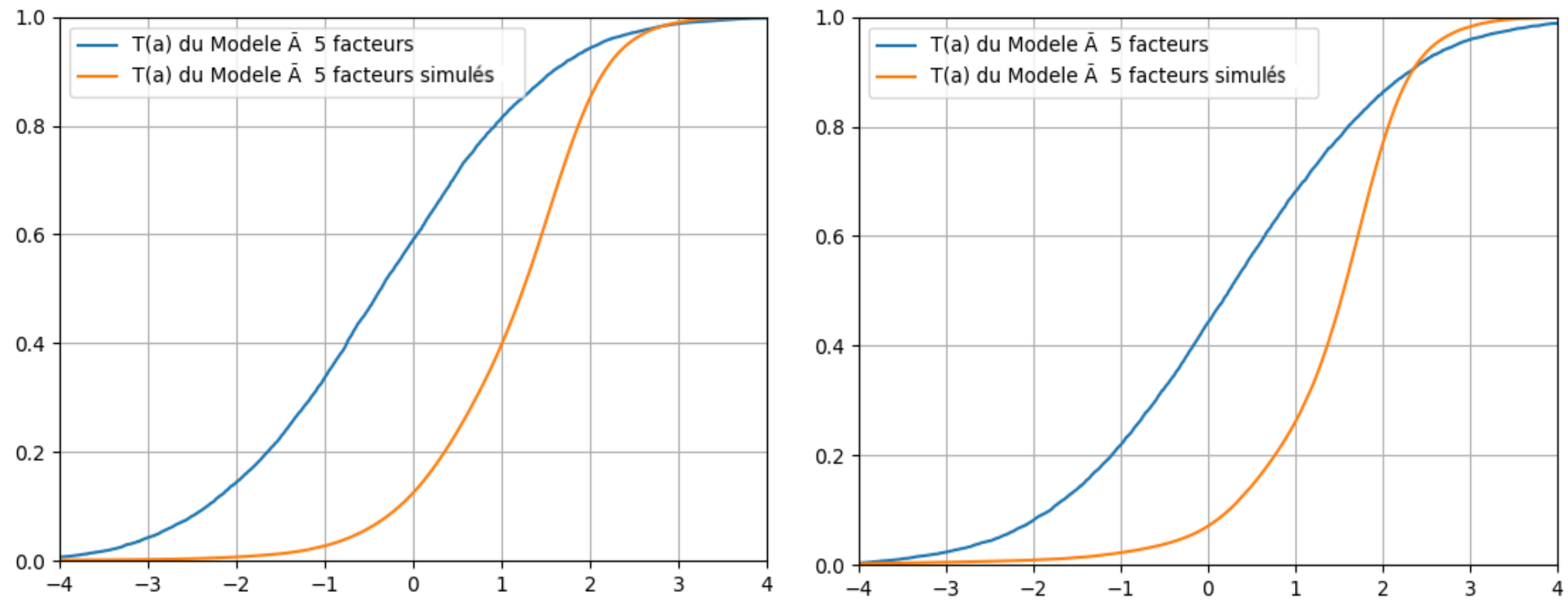

(a) Result for 5 Million AuM Funds (2005-2019) (b) Result for 1 Billion AuM Funds (2005-2019)

Figure 10: Comparison of the cumulative distribution functions for the two group of funds

# 4 Conclusion

Our two questions before starting to work on this paper concerned the effects of new performance attribution models in the asset management industry and how to assess the accuracy of models. Thus we were able to see and show that the extension of old models is far from futile, and can shed new light or correct existing biases. We have also been able to highlight the impact that these models have had on the asset management industry and performance of fund managers. Although the 5-factor model helps to mitigate the observed differences between small and large funds, it is ultimately the large funds that have been best able to adapt to this new environment. During the course of our study we were able to highlight an increase in the net benchmark-adjusted results for all groups, but contrasted with an implementation based on fund size.

This study will also have made it possible to complete the study previously carried out by Fama French, but does not completely change the conclusion. In fact, even if there seem to be more managers talented enough to create a positive net benchmark adjusted performance, it appears that they are still a minority. This conclusion is disappointing for investors and partly explains their choice of ETFs despite diversified funds. However, choosing to invest in an actively managed diversified fund can be a wise choice, provided that the manager is well chosen. It was also established during this study that choosing a large fund is preferable as it increases the expectation of good management. However, this is not a guarantee of quality, as the overwhelming majority of these large funds also underperform.

# 5 Appendix

## 5.1 Tables

### 5.1.1 Centiles Period 1984-2006

| Pct | 5 Million | | | 250 Million | | | 1 Billion | | |
|---|---|---|---|---|---|---|---|---|---|
| | Sim | Act | %<Act | Sim | Act | %<Act | Sim | Act | %<Act |
| 3-Factor Net Returns | | | | | | | | | |
| 1 | −2.50 | −3.87 | 0.08 | −2.45 | −3.87 | 0.10 | −2.39 | −4.39 | 0.01 |
| 2 | −2.17 | −3.42 | 0.06 | −2.13 | −3.38 | 0.13 | −2.09 | −3.55 | 0.09 |
| 3 | −1.97 | −3.15 | 0.07 | −1.94 | −3.15 | 0.12 | −1.91 | −3.36 | 0.07 |
| 4 | −1.83 | −2.99 | 0.06 | −1.80 | −3.04 | 0.10 | −1.78 | −3.16 | 0.07 |
| 5 | −1.71 | −2.84 | 0.08 | −1.69 | −2.91 | 0.10 | −1.67 | −2.99 | 0.10 |
| 10 | −1.32 | −2.34 | 0.05 | −1.31 | −2.37 | 0.10 | −1.30 | −2.53 | 0.08 |
| 20 | −0.87 | −1.74 | 0.03 | −0.86 | −1.87 | 0.04 | −0.86 | −1.98 | 0.03 |
| 30 | −0.54 | −1.27 | 0.06 | −0.54 | −1.41 | 0.06 | −0.54 | −1.59 | 0.02 |
| 40 | −0.26 | −0.92 | 0.05 | −0.27 | −1.03 | 0.07 | −0.27 | −1.19 | 0.02 |
| 50 | −0.01 | −0.62 | 0.04 | −0.01 | −0.71 | 0.06 | −0.01 | −0.82 | 0.03 |
| 60 | 0.25 | −0.29 | 0.11 | 0.25 | −0.39 | 0.19 | 0.24 | −0.51 | 0.05 |
| 70 | 0.52 | 0.08 | 0.51 | 0.52 | −0.08 | 0.25 | 0.52 | −0.20 | 0.08 |
| 80 | 0.85 | 0.50 | 3.20 | 0.84 | 0.37 | 1.68 | 0.84 | 0.25 | 0.85 |
| 90 | 1.30 | 1.01 | 8.17 | 1.29 | 0.89 | 5.19 | 1.28 | 0.82 | 4.81 |
| 95 | 1.68 | 1.54 | 30.55 | 1.66 | 1.36 | 14.17 | 1.64 | 1.34 | 17.73 |
| 96 | 1.80 | 1.71 | 40.06 | 1.76 | 1.49 | 17.24 | 1.74 | 1.52 | 26.33 |
| 97 | 1.94 | 1.91 | 49.35 | 1.90 | 1.69 | 25.92 | 1.87 | 1.79 | 42.86 |
| 98 | 2.13 | 2.17 | 58.70 | 2.08 | 1.90 | 30.43 | 2.04 | 2.02 | 50.07 |
| 99 | 2.45 | 2.47 | 57.42 | 2.36 | 2.29 | 43.92 | 2.31 | 2.40 | 63.11 |
| 4-Factor Net Returns | | | | | | | | | |
| 1 | −2.55 | −3.94 | 0.04 | −2.47 | −3.94 | 0.08 | −2.40 | −4.22 | 0.01 |
| 2 | −2.20 | −3.43 | 0.04 | −2.14 | −3.43 | 0.09 | −2.09 | −3.48 | 0.08 |
| 3 | −2.00 | −3.08 | 0.13 | −1.95 | −3.07 | 0.25 | −1.91 | −3.11 | 0.23 |
| 4 | −1.85 | −2.88 | 0.13 | −1.80 | −2.88 | 0.22 | −1.77 | −2.95 | 0.21 |
| 5 | −1.73 | −2.74 | 0.12 | −1.69 | −2.78 | 0.18 | −1.66 | −2.86 | 0.14 |
| 10 | −1.33 | −2.23 | 0.14 | −1.30 | −2.34 | 0.14 | −1.29 | −2.48 | 0.07 |
| 20 | −0.86 | −1.67 | 0.10 | −0.85 | −1.80 | 0.11 | −0.84 | −1.96 | 0.05 |
| 30 | −0.53 | −1.25 | 0.12 | −0.52 | −1.39 | 0.10 | −0.52 | −1.54 | 0.04 |
| 40 | −0.25 | −0.88 | 0.21 | −0.25 | −1.04 | 0.14 | −0.25 | −1.23 | 0.05 |
| 50 | 0.01 | −0.60 | 0.18 | 0.01 | −0.76 | 0.11 | 0.01 | −0.87 | 0.07 |
| 60 | 0.26 | −0.29 | 0.25 | 0.27 | −0.42 | 0.29 | 0.26 | −0.49 | 0.19 |
| 70 | 0.54 | 0.02 | 0.37 | 0.54 | −0.13 | 0.24 | 0.54 | −0.18 | 0.24 |
| 80 | 0.87 | 0.44 | 1.76 | 0.86 | 0.27 | 0.72 | 0.86 | 0.17 | 0.45 |
| 90 | 1.33 | 1.04 | 10.62 | 1.31 | 0.86 | 4.40 | 1.30 | 0.86 | 7.07 |
| 95 | 1.72 | 1.53 | 23.82 | 1.69 | 1.37 | 14.35 | 1.67 | 1.31 | 14.13 |
| 96 | 1.84 | 1.67 | 28.21 | 1.80 | 1.51 | 18.23 | 1.78 | 1.45 | 17.16 |
| 97 | 1.99 | 1.84 | 31.30 | 1.94 | 1.65 | 18.62 | 1.91 | 1.57 | 17.05 |
| 98 | 2.19 | 2.09 | 39.12 | 2.12 | 1.79 | 15.57 | 2.08 | 1.76 | 18.86 |
| 99 | 2.52 | 2.40 | 36.96 | 2.42 | 2.22 | 29.88 | 2.36 | 2.26 | 42.00 |

(a) Results Fama and French (2010)

| Pct | 5 Millions | | | 250 Millions | | | 1 Milliard | | |
|---|---|---|---|---|---|---|---|---|---|
| | Sim | Act | %<Act | Sim | Act | %<Act | Sim | Act | %<Act |
| 3 Facteurs | | *2363 Fonds* | | | *583 Fonds* | | | *382 Fonds* | |
| 1 | -2,65 | -4,50 | 0,26 | -2,98 | -3,69 | 0,72 | -1,39 | -3,16 | 0,23 |
| 2 | -1,84 | -3,68 | 0,43 | -1,60 | -3,18 | 0,91 | -1,12 | -2,76 | 0,25 |
| 3 | -1,46 | -3,44 | 0,52 | -1,23 | -2,91 | 1,03 | -0,95 | -2,30 | 0,27 |
| 4 | -1,23 | -3,28 | 0,58 | -1,03 | -2,65 | 1,15 | -0,79 | -2,16 | 0,28 |
| 5 | -1,05 | -3,13 | 0,66 | -0,89 | -2,51 | 1,22 | -0,63 | -2,05 | 0,29 |
| 10 | -0,51 | -2,53 | 1,11 | -0,43 | -1,84 | 1,66 | -0,03 | -1,48 | 0,78 |
| 20 | 0,06 | -1,83 | 2,03 | 0,17 | -1,20 | 3,11 | 0,54 | -0,87 | 3,49 |
| 30 | 0,42 | -1,37 | 3,36 | 0,54 | -0,73 | 6,50 | 0,86 | -0,36 | 6,87 |
| 40 | 0,70 | -1,01 | 5,31 | 0,83 | -0,23 | 12,79 | 1,10 | 0,13 | 11,98 |
| 50 | 0,94 | -0,63 | 8,65 | 1,08 | 0,18 | 20,33 | 1,33 | 0,49 | 18,62 |
| 60 | 1,18 | -0,24 | 13,96 | 1,33 | 0,56 | 30,54 | 1,56 | 0,86 | 30,28 |
| 70 | 1,44 | 0,21 | 23,59 | 1,60 | 0,95 | 44,86 | 1,85 | 1,15 | 42,27 |
| 80 | 1,74 | 0,73 | 41,34 | 1,93 | 1,40 | 62,88 | 2,20 | 1,55 | 59,41 |
| 90 | 2,18 | 1,34 | 66,21 | 2,39 | 1,96 | 80,73 | 2,67 | 2,19 | 79,71 |
| 95 | 2,53 | 1,90 | 84,06 | 2,73 | 2,40 | 90,13 | 2,99 | 2,66 | 89,89 |
| 96 | 2,62 | 2,02 | 86,93 | 2,81 | 2,54 | 92,44 | 3,07 | 2,90 | 93,77 |
| 97 | 2,74 | 2,21 | 90,56 | 2,91 | 2,65 | 93,90 | 3,15 | 2,95 | 94,51 |
| 98 | 2,89 | 2,37 | 93,13 | 3,02 | 2,84 | 96,30 | 3,25 | 3,01 | 95,25 |
| 99 | 3,12 | 2,66 | 96,35 | 3,19 | 3,18 | 98,95 | 3,40 | 3,39 | 98,96 |
| 5 Facteurs | | *2363 Fonds* | | | *583 Fonds* | | | *382 Fonds* | |
| 1 | -2,50 | -4,53 | 0,26 | -2,80 | -4,22 | 0,56 | -1,42 | -3,86 | 0,15 |
| 2 | -1,76 | -4,05 | 0,34 | -1,64 | -3,75 | 0,72 | -1,08 | -3,40 | 0,20 |
| 3 | -1,43 | -3,77 | 0,41 | -1,34 | -3,46 | 0,81 | -0,87 | -3,20 | 0,22 |
| 4 | -1,23 | -3,58 | 0,46 | -1,17 | -3,12 | 0,90 | -0,71 | -3,16 | 0,22 |
| 5 | -1,07 | -3,34 | 0,54 | -1,03 | -2,94 | 0,95 | -0,58 | -3,00 | 0,23 |
| 10 | -0,60 | -2,64 | 0,90 | -0,56 | -2,40 | 1,17 | -0,12 | -2,22 | 0,29 |
| 20 | -0,07 | -1,88 | 1,78 | 0,03 | -1,49 | 2,41 | 0,42 | -1,34 | 1,18 |
| 30 | 0,30 | -1,41 | 3,08 | 0,42 | -0,78 | 7,39 | 0,76 | -0,86 | 3,04 |
| 40 | 0,59 | -1,00 | 5,54 | 0,72 | -0,23 | 15,00 | 1,05 | -0,18 | 9,28 |
| 50 | 0,86 | -0,57 | 10,50 | 0,98 | 0,32 | 27,16 | 1,31 | 0,36 | 18,48 |
| 60 | 1,12 | -0,13 | 18,44 | 1,25 | 0,72 | 39,93 | 1,60 | 0,72 | 28,48 |
| 70 | 1,40 | 0,35 | 31,60 | 1,55 | 1,12 | 55,18 | 1,93 | 1,21 | 46,24 |
| 80 | 1,73 | 0,83 | 48,99 | 1,92 | 1,56 | 70,31 | 2,34 | 1,60 | 59,99 |
| 90 | 2,21 | 1,47 | 72,42 | 2,51 | 2,14 | 84,32 | 2,90 | 2,29 | 78,92 |
| 95 | 2,63 | 1,97 | 85,70 | 2,96 | 2,49 | 89,79 | 3,28 | 2,93 | 90,56 |
| 96 | 2,75 | 2,11 | 88,29 | 3,08 | 2,64 | 91,58 | 3,38 | 3,03 | 91,94 |
| 97 | 2,89 | 2,25 | 90,61 | 3,21 | 2,78 | 93,21 | 3,49 | 3,13 | 93,25 |
| 98 | 3,08 | 2,47 | 93,48 | 3,37 | 3,12 | 96,31 | 3,64 | 3,30 | 95,20 |
| 99 | 3,36 | 2,79 | 96,32 | 3,60 | 3,64 | 99,13 | 3,84 | 3,57 | 97,57 |

(b) Results Observed

Figure 11: Centile comparison on 1984-2006 period

### 5.1.2 Centiles Period 1990-2005

| Période 1990-2005 | | | | | | | | | |
|---|---|---|---|---|---|---|---|---|---|
| | 5 Millions | | | 250 Millions | | | 1 Milliard | | |
| Pct | Sim | Act | %<Act | Sim | Act | %<Act | Sim | Act | %<Act |
| 3 Facteurs | *2735 Fonds* | | | *652 Fonds* | | | *427 Fonds* | | |
| 1 | -1,33 | -4,22 | 0,18 | -1,02 | -3,30 | 0,15 | -1,03 | -2,80 | 0,00 |
| 2 | -0,95 | -3,67 | 0,19 | -0,67 | -3,15 | 0,15 | -0,76 | -2,20 | 0,01 |
| 3 | -0,73 | -3,39 | 0,19 | -0,44 | -3,01 | 0,16 | -0,49 | -2,06 | 0,02 |
| 4 | -0,58 | -3,19 | 0,19 | -0,27 | -2,59 | 0,16 | -0,26 | -1,95 | 0,03 |
| 5 | -0,45 | -3,03 | 0,20 | -0,14 | -2,41 | 0,17 | -0,09 | -1,88 | 0,04 |
| 10 | -0,03 | -2,37 | 0,26 | 0,24 | -1,77 | 0,28 | 0,36 | -1,46 | 0,24 |
| 20 | 0,43 | -1,80 | 0,47 | 0,62 | -1,14 | 0,78 | 0,76 | -0,87 | 1,55 |
| 30 | 0,73 | -1,37 | 0,93 | 0,87 | -0,71 | 1,85 | 1,04 | -0,49 | 3,01 |
| 40 | 0,98 | -1,01 | 1,78 | 1,09 | -0,38 | 3,36 | 1,28 | -0,11 | 4,87 |
| 50 | 1,21 | -0,66 | 3,42 | 1,31 | -0,03 | 6,02 | 1,50 | 0,37 | 10,28 |
| 60 | 1,44 | -0,27 | 6,76 | 1,53 | 0,44 | 14,44 | 1,71 | 0,75 | 19,62 |
| 70 | 1,68 | 0,11 | 12,44 | 1,76 | 0,80 | 26,94 | 1,94 | 1,07 | 31,32 |
| 80 | 1,96 | 0,60 | 25,27 | 2,03 | 1,16 | 42,88 | 2,19 | 1,50 | 50,08 |
| 90 | 2,35 | 1,16 | 47,67 | 2,38 | 1,67 | 66,30 | 2,48 | 1,96 | 70,93 |
| 95 | 2,66 | 1,69 | 70,26 | 2,63 | 2,07 | 81,51 | 2,68 | 2,56 | 92,19 |
| 96 | 2,75 | 1,85 | 76,47 | 2,70 | 2,22 | 85,89 | 2,73 | 2,72 | 95,79 |
| 97 | 2,86 | 2,06 | 83,13 | 2,78 | 2,36 | 89,53 | 2,79 | 2,95 | 98,63 |
| 98 | 3,00 | 2,26 | 88,05 | 2,89 | 2,58 | 94,17 | 2,88 | 3,05 | 99,19 |
| 99 | 3,23 | 2,56 | 93,72 | 3,07 | 2,98 | 98,58 | 3,01 | 3,30 | 99,81 |
| 5 Facteurs | *2735 Fonds* | | | *652 Fonds* | | | *427 Fonds* | | |
| 1 | -1,37 | -4,46 | 0,18 | -1,14 | -4,06 | 0,15 | -0,96 | -3,84 | 0,00 |
| 2 | -1,01 | -4,05 | 0,18 | -0,80 | -3,72 | 0,15 | -0,65 | -3,39 | 0,00 |
| 3 | -0,80 | -3,73 | 0,19 | -0,60 | -3,42 | 0,15 | -0,46 | -3,14 | 0,00 |
| 4 | -0,66 | -3,44 | 0,19 | -0,45 | -3,10 | 0,16 | -0,31 | -3,03 | 0,00 |
| 5 | -0,55 | -3,22 | 0,20 | -0,34 | -2,88 | 0,16 | -0,19 | -2,79 | 0,00 |
| 10 | -0,17 | -2,60 | 0,24 | 0,04 | -2,23 | 0,20 | 0,21 | -2,17 | 0,03 |
| 20 | 0,28 | -1,86 | 0,46 | 0,46 | -1,53 | 0,47 | 0,65 | -1,53 | 0,26 |
| 30 | 0,59 | -1,40 | 0,94 | 0,75 | -0,97 | 1,41 | 0,96 | -0,94 | 1,05 |
| 40 | 0,86 | -0,99 | 2,08 | 1,00 | -0,52 | 3,47 | 1,23 | -0,52 | 2,62 |
| 50 | 1,11 | -0,60 | 4,49 | 1,23 | 0,17 | 12,44 | 1,49 | 0,03 | 7,37 |
| 60 | 1,36 | -0,18 | 9,75 | 1,48 | 0,49 | 20,78 | 1,77 | 0,57 | 17,85 |
| 70 | 1,63 | 0,26 | 19,47 | 1,76 | 0,87 | 34,62 | 2,06 | 1,03 | 32,64 |
| 80 | 1,95 | 0,70 | 34,17 | 2,10 | 1,20 | 48,42 | 2,40 | 1,49 | 49,76 |
| 90 | 2,39 | 1,32 | 58,48 | 2,58 | 1,87 | 73,42 | 2,81 | 2,17 | 73,38 |
| 95 | 2,76 | 1,82 | 76,08 | 2,95 | 2,32 | 85,15 | 3,11 | 2,60 | 85,18 |
| 96 | 2,86 | 1,95 | 79,98 | 3,04 | 2,46 | 87,95 | 3,19 | 2,72 | 87,96 |
| 97 | 2,99 | 2,10 | 83,79 | 3,16 | 2,69 | 91,77 | 3,29 | 2,78 | 89,39 |
| 98 | 3,15 | 2,23 | 86,78 | 3,32 | 2,83 | 93,64 | 3,41 | 3,12 | 95,15 |
| 99 | 3,40 | 2,74 | 94,78 | 3,55 | 3,05 | 96,09 | 3,59 | 3,73 | 99,43 |

Figure 12: Results for 3-factor 5-factor models (1990-2005)

### 5.1.3 Centiles Period 2005-2019

| Période2005-2020 | | | | | | | | | |
|---|---|---|---|---|---|---|---|---|---|
| | 5 Millions | | | 250 Millions | | | 1 Milliard | | |
| Pct | Sim | Act | %<Act | Sim | Act | %<Act | Sim | Act | %<Act |
| 3 Facteurs | *6252 Fonds* | | | *1843 Fonds* | | | *1157 Fonds* | | |
| 1 | -1,71 | -3,85 | 0,16 | -1,69 | -3,46 | 0,14 | -1,84 | -3,19 | 0,43 |
| 2 | -1,18 | -3,42 | 0,19 | -1,11 | -3,16 | 0,24 | -1,17 | -2,84 | 0,53 |
| 3 | -0,92 | -3,19 | 0,23 | -0,78 | -2,91 | 0,35 | -0,80 | -2,57 | 0,62 |
| 4 | -0,72 | -3,01 | 0,27 | -0,52 | -2,70 | 0,45 | -0,45 | -2,43 | 0,68 |
| 5 | -0,57 | -2,86 | 0,30 | -0,30 | -2,60 | 0,49 | -0,18 | -2,29 | 0,74 |
| 10 | -0,01 | -2,42 | 0,44 | 0,33 | -2,09 | 0,69 | 0,42 | -1,82 | 1,01 |
| 20 | 0,51 | -1,79 | 0,92 | 0,80 | -1,42 | 1,38 | 0,92 | -1,19 | 1,94 |
| 30 | 0,84 | -1,31 | 1,62 | 1,16 | -0,90 | 2,61 | 1,31 | -0,70 | 3,27 |
| 40 | 1,15 | -0,89 | 3,13 | 1,41 | -0,46 | 4,26 | 1,55 | -0,26 | 4,67 |
| 50 | 1,40 | -0,48 | 5,66 | 1,59 | -0,09 | 6,14 | 1,71 | 0,16 | 7,03 |
| 60 | 1,59 | -0,10 | 8,95 | 1,73 | 0,31 | 9,73 | 1,84 | 0,60 | 13,25 |
| 70 | 1,77 | 0,30 | 14,99 | 1,87 | 0,73 | 18,20 | 1,96 | 1,06 | 23,09 |
| 80 | 1,94 | 0,81 | 29,07 | 2,01 | 1,22 | 32,05 | 2,09 | 1,54 | 39,44 |
| 90 | 2,18 | 1,50 | 55,06 | 2,22 | 1,89 | 71,32 | 2,31 | 2,19 | 85,72 |
| 95 | 2,44 | 2,14 | 88,79 | 2,46 | 2,55 | 95,99 | 2,59 | 2,74 | 96,36 |
| 96 | 2,53 | 2,32 | 93,27 | 2,55 | 2,69 | 97,14 | 2,69 | 2,85 | 97,14 |
| 97 | 2,65 | 2,52 | 95,86 | 2,67 | 2,83 | 97,99 | 2,83 | 3,08 | 98,51 |
| 98 | 2,80 | 2,79 | 97,96 | 2,83 | 2,98 | 98,78 | 2,99 | 3,53 | 99,74 |
| 99 | 3,01 | 3,30 | 99,67 | 3,04 | 3,53 | 99,87 | 3,19 | 4,52 | 99,91 |
| 5 Facteurs | *6252 Fonds* | | | *1843 Fonds* | | | *1157 Fonds* | | |
| 1 | -1,72 | -3,62 | 0,17 | -1,67 | -3,30 | 0,19 | -1,84 | -2,91 | 0,52 |
| 2 | -1,21 | -3,25 | 0,21 | -1,05 | -3,02 | 0,29 | -1,10 | -2,71 | 0,59 |
| 3 | -0,94 | -3,01 | 0,25 | -0,73 | -2,76 | 0,40 | -0,72 | -2,47 | 0,68 |
| 4 | -0,76 | -2,85 | 0,28 | -0,51 | -2,59 | 0,47 | -0,45 | -2,30 | 0,75 |
| 5 | -0,62 | -2,71 | 0,32 | -0,34 | -2,48 | 0,51 | -0,26 | -2,20 | 0,79 |
| 10 | -0,16 | -2,25 | 0,53 | 0,14 | -1,98 | 0,75 | 0,24 | -1,71 | 1,11 |
| 20 | 0,35 | -1,66 | 1,08 | 0,64 | -1,29 | 1,50 | 0,76 | -1,06 | 2,11 |
| 30 | 0,71 | -1,20 | 2,01 | 0,98 | -0,78 | 2,83 | 1,12 | -0,62 | 3,33 |
| 40 | 1,01 | -0,77 | 3,94 | 1,23 | -0,39 | 4,68 | 1,36 | -0,22 | 5,25 |
| 50 | 1,25 | -0,40 | 7,03 | 1,43 | -0,03 | 7,82 | 1,55 | 0,17 | 9,03 |
| 60 | 1,45 | -0,02 | 12,21 | 1,60 | 0,40 | 14,56 | 1,72 | 0,64 | 17,31 |
| 70 | 1,65 | 0,42 | 21,55 | 1,77 | 0,82 | 24,86 | 1,88 | 1,07 | 28,36 |
| 80 | 1,87 | 0,89 | 35,86 | 1,96 | 1,29 | 42,71 | 2,06 | 1,50 | 47,27 |
| 90 | 2,16 | 1,56 | 65,52 | 2,23 | 1,92 | 78,40 | 2,33 | 2,26 | 87,86 |
| 95 | 2,43 | 2,12 | 88,79 | 2,47 | 2,49 | 95,19 | 2,60 | 2,74 | 96,50 |
| 96 | 2,51 | 2,27 | 92,39 | 2,55 | 2,62 | 96,75 | 2,68 | 2,90 | 97,69 |
| 97 | 2,62 | 2,47 | 95,57 | 2,65 | 2,81 | 98,09 | 2,80 | 3,20 | 98,99 |
| 98 | 2,76 | 2,77 | 98,04 | 2,79 | 3,06 | 99,15 | 2,95 | 3,60 | 99,71 |
| 99 | 3,00 | 3,25 | 99,55 | 3,01 | 3,45 | 99,80 | 3,20 | 4,43 | 99,91 |

Figure 13: Results for 3 and 5 factors models (2005-2019)

## 5.2 Graphics

### 5.2.1 Graphics 3 Factors 1984-2006

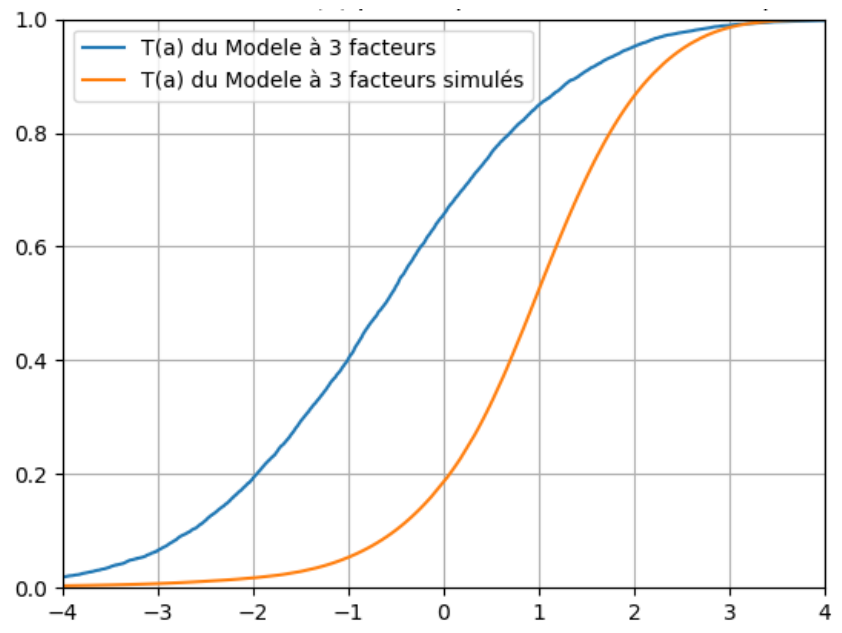

Figure 14: Cumulative density function for 3 factors model 1984-2006 Group 5 millions AUM

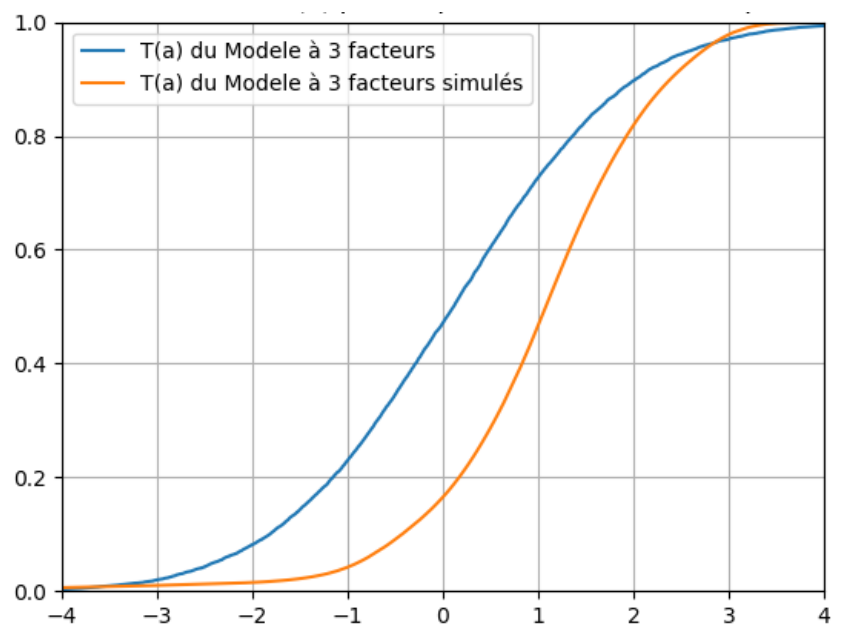

Figure 15: Cumulative density function for 3 factors model 1984-2006 Group 250 millions AUM

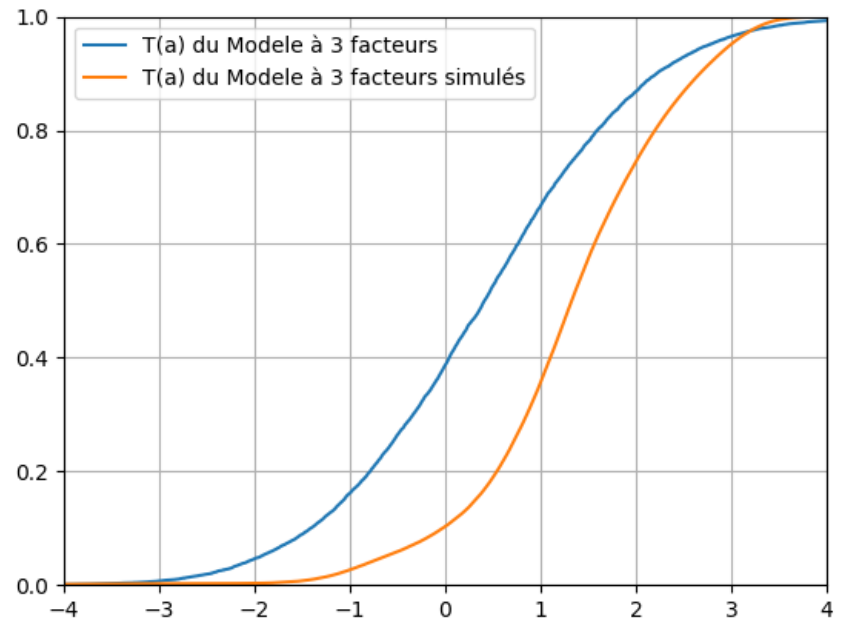

Figure 16: Cumulative density function for 3 factors model 1984-2006 Group 1000 millions AUM

### 5.2.2 Graphics 3 Factors 1990-2005

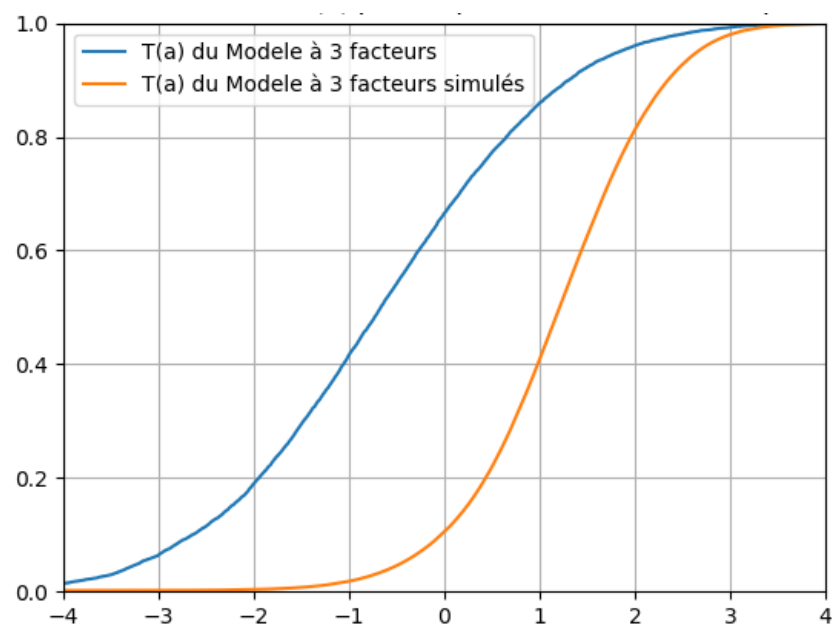

Figure 17: Cumulative density function for 3 factors model 1990-2005 Group 5 millions AUM

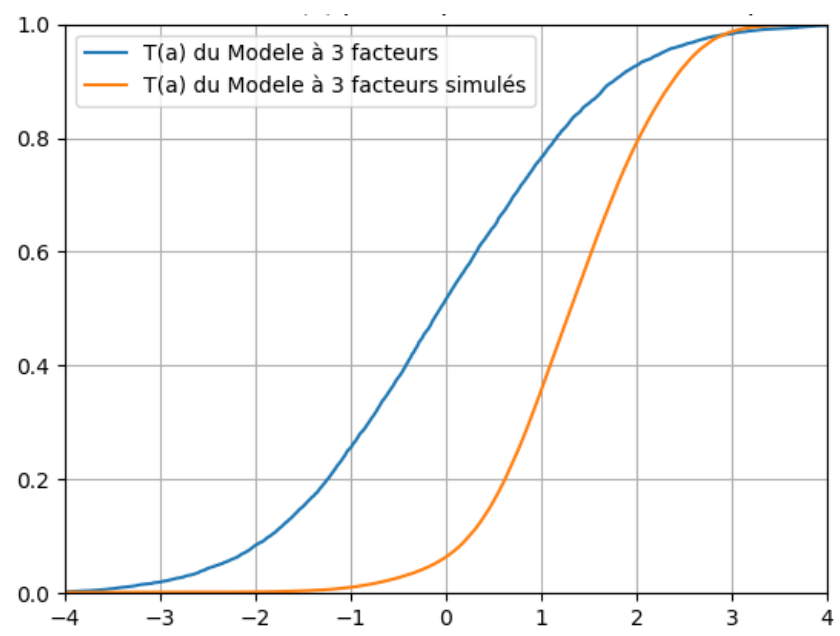

Figure 18: Cumulative density function for 3 factors model 1990-2005 Group 250 millions AUM

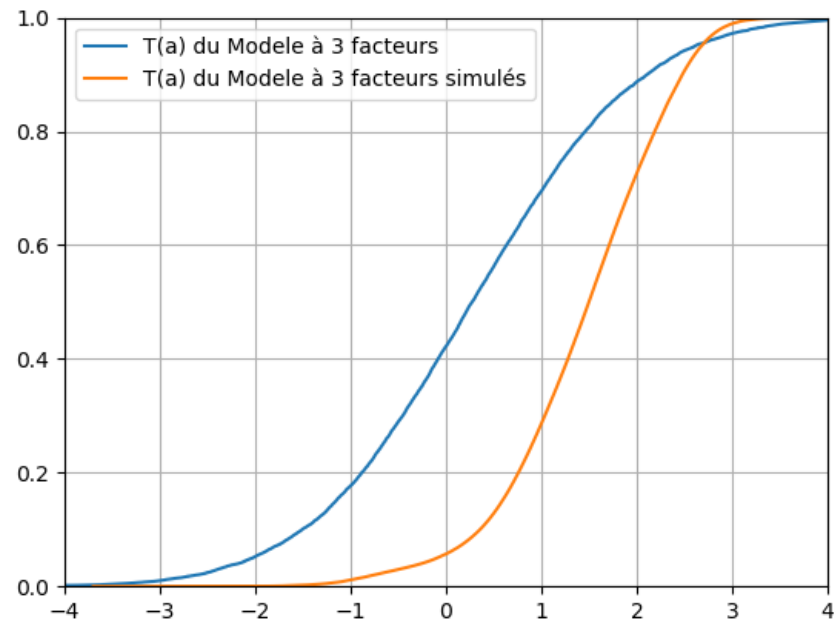

Figure 19: Cumulative density function for 3 factors model 1990-2005 Group 1000 millions AUM

### 5.2.3 Graphics 3 Factors 2005-2019

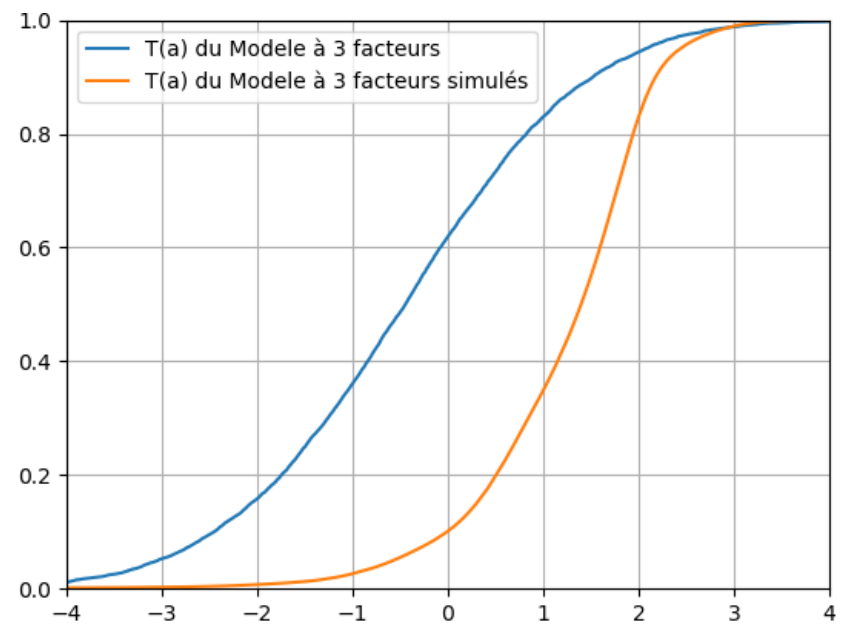

Figure 20: Cumulative density function for 3 factors model 2005-2019 Group 5 millions AUM

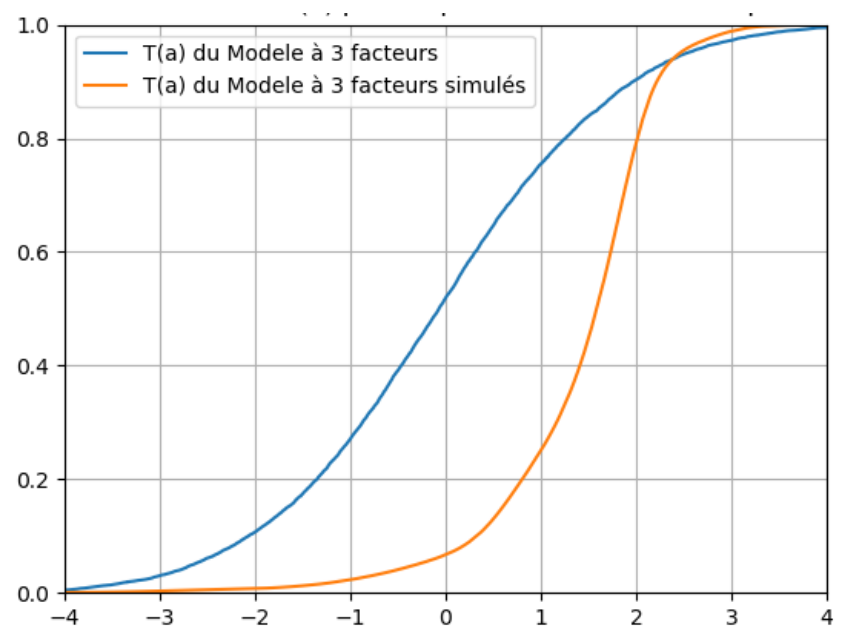

Figure 21: Cumulative density function for 3 factors model 2005-2019 Group 250 millions AUM

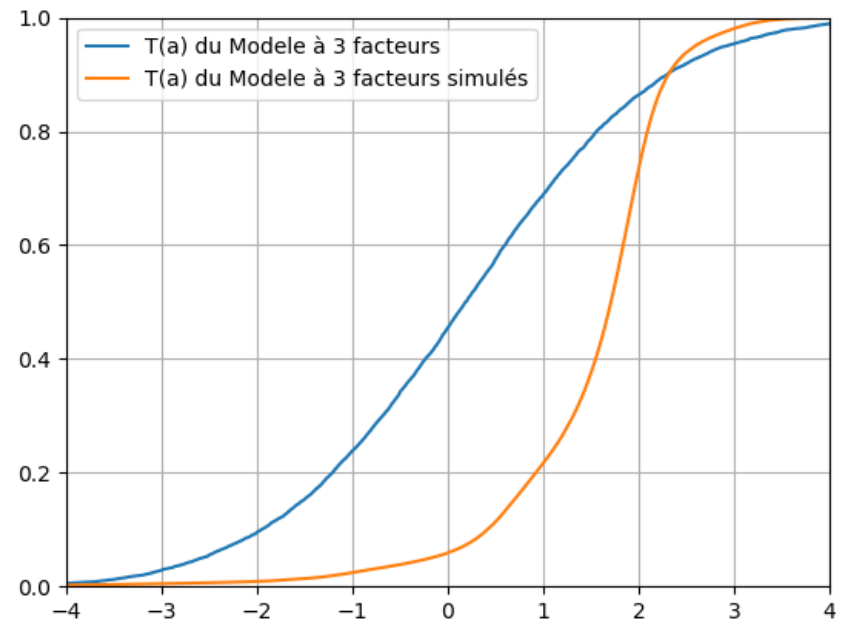

Figure 22: Cumulative density function for 3 factors model 2005-2019 Group 1000 millions AUM

### 5.2.4 Graphics 5 Factors 1984-2006

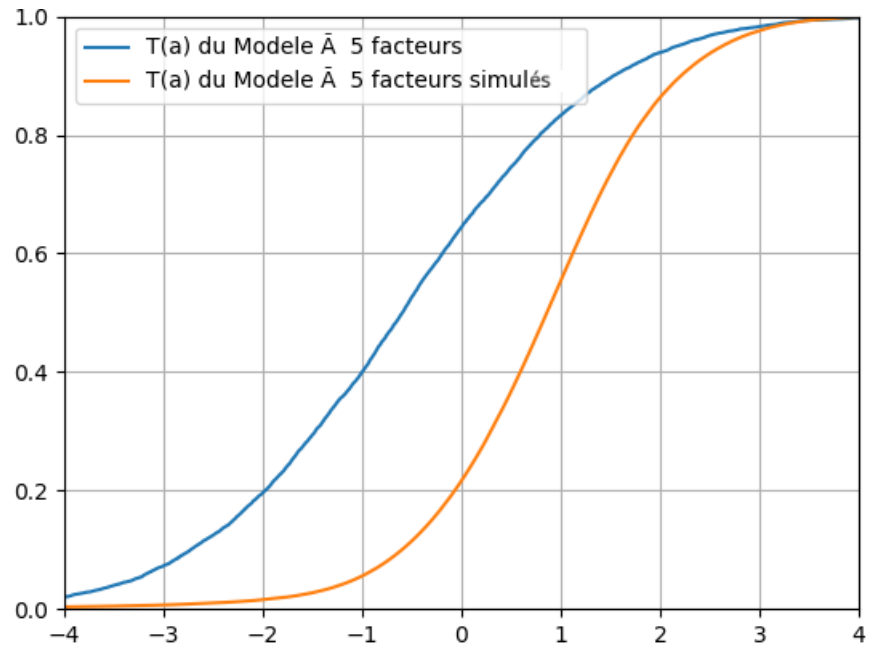

Figure 23: Cumulative density function for 5 factors model 1984-2006 Group 5 millions AUM

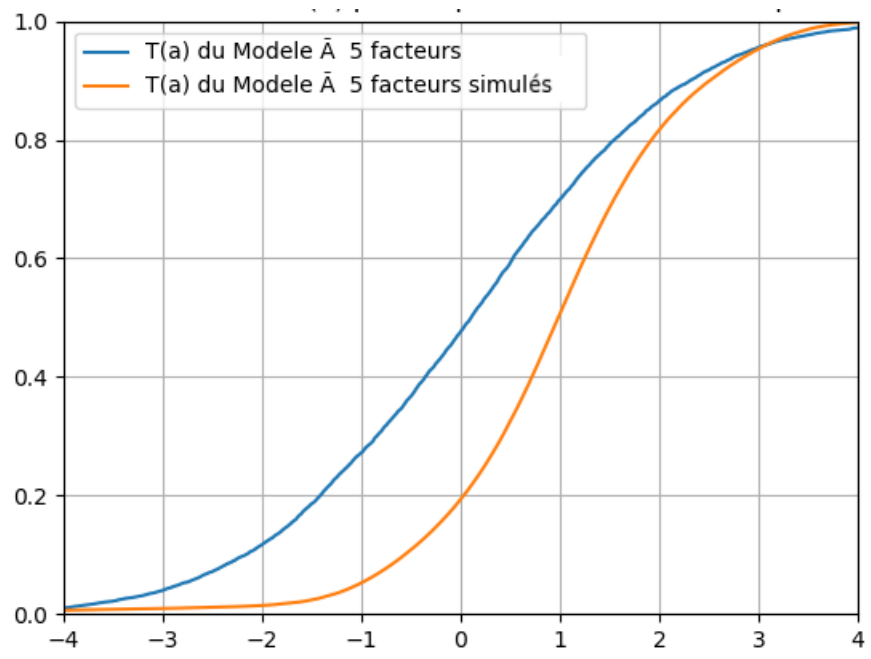

Figure 24: Cumulative density function for 5 factors model 1984-2006 Group 250 millions AUM

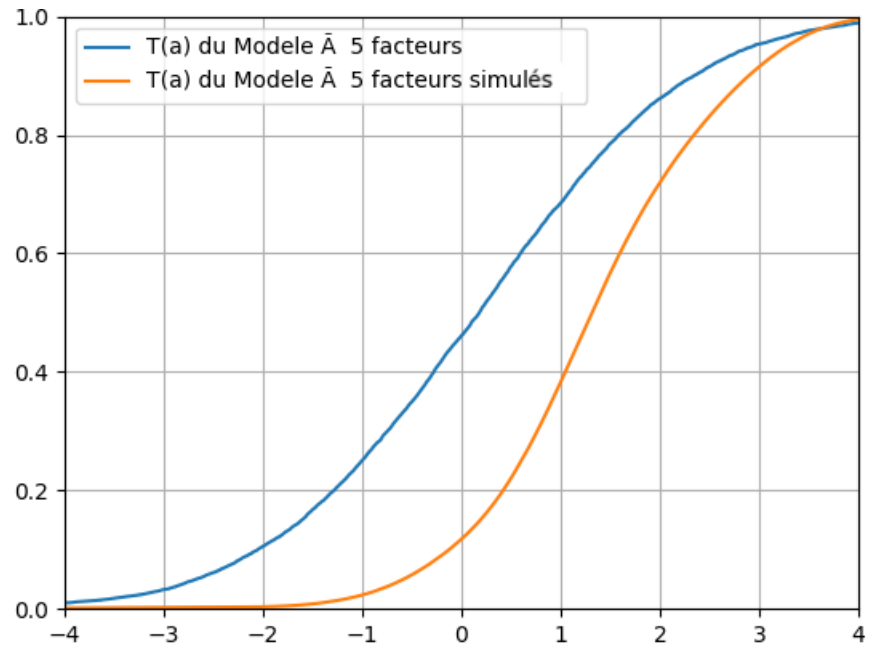

Figure 25: Cumulative density function for 5 factors model 1984-2006 Group 1000 millions AUM

### 5.2.5 Graphics 5 Factors 1990-2005

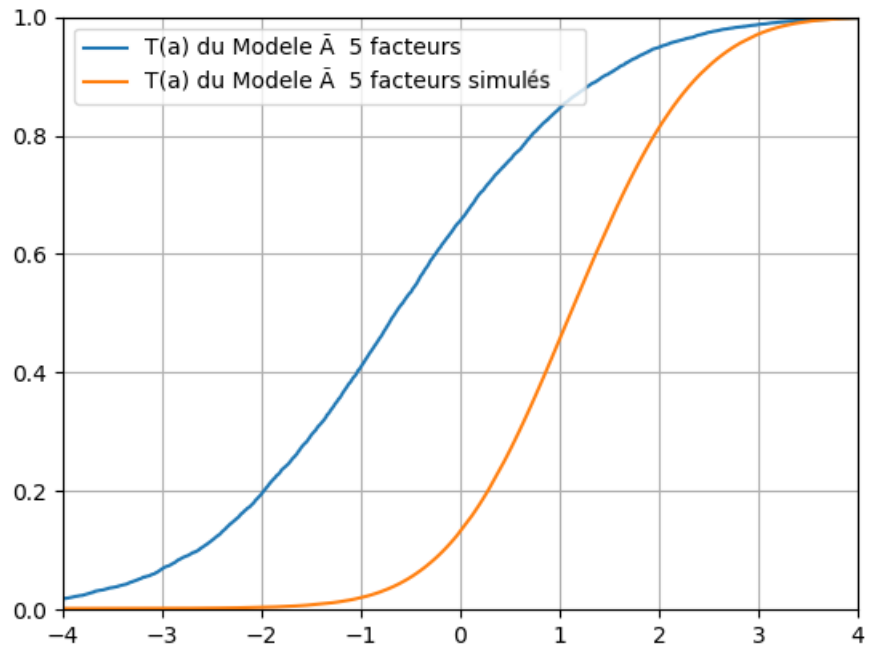

Figure 26: Cumulative density function for 5 factors model 1990-2005 Group 5 millions AUM

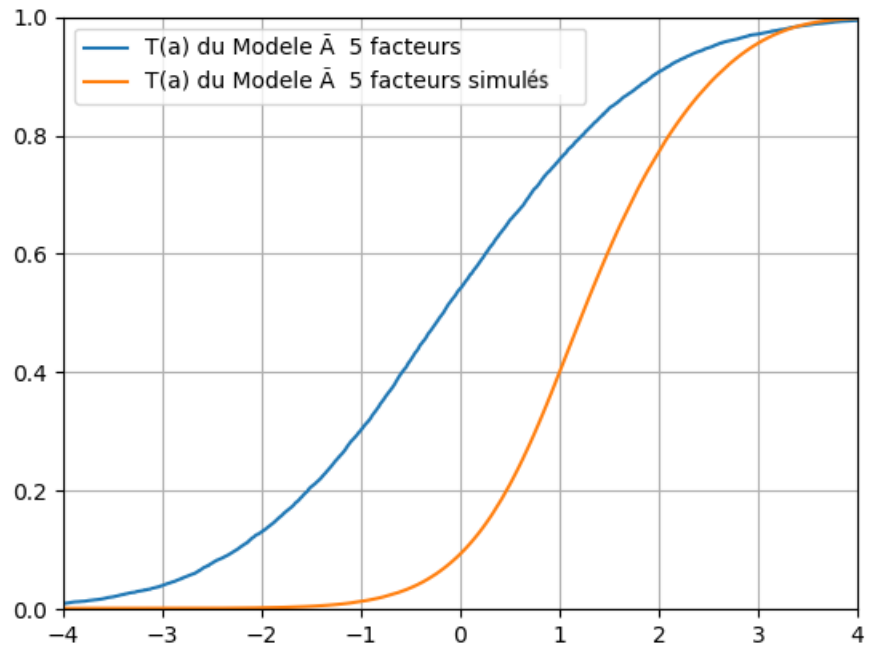

Figure 27: Cumulative density function for 5 factors model 1990-2005 Group 250 millions AUM

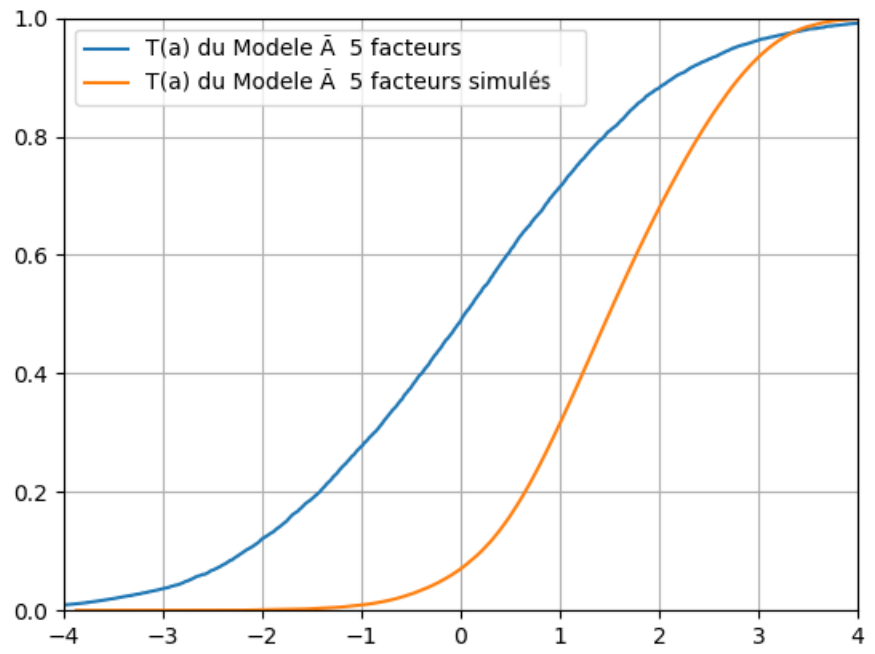

Figure 28: Cumulative density function for 5 factors model 1990-2005 Group 1000 millions AUM

### 5.2.6 Graphics 5 Factors 2005-2019

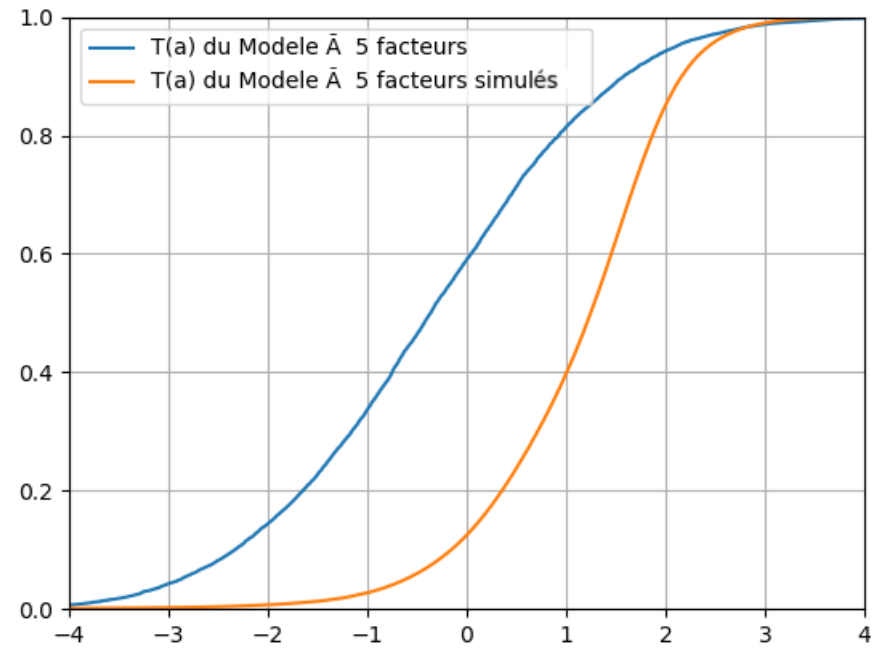

Figure 29: Cumulative density function for 5 factors model 2005-2019 Group 5 millions AUM

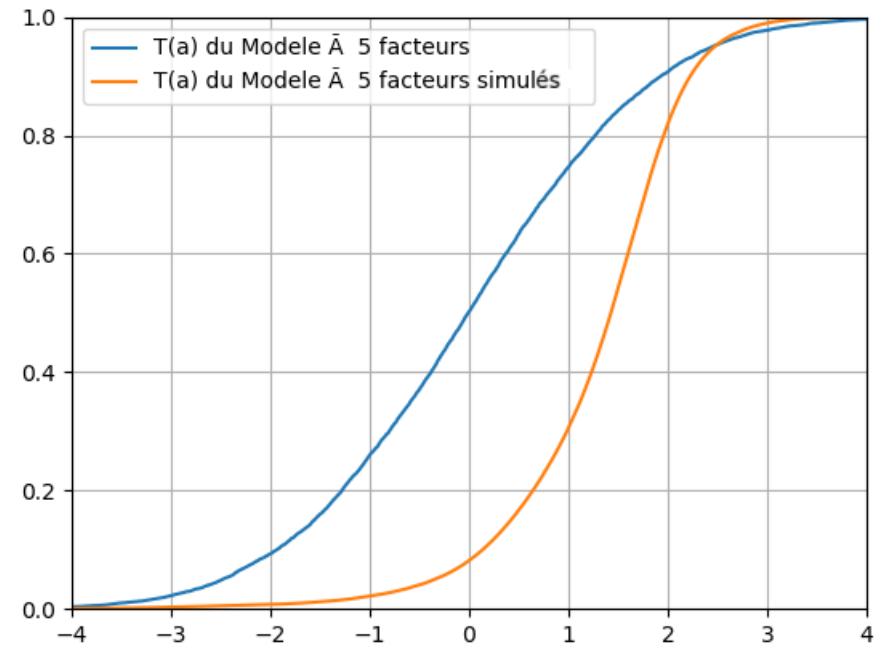

Figure 30: Cumulative density function for 5 factors model 2005-2019 Group 250 millions AUM

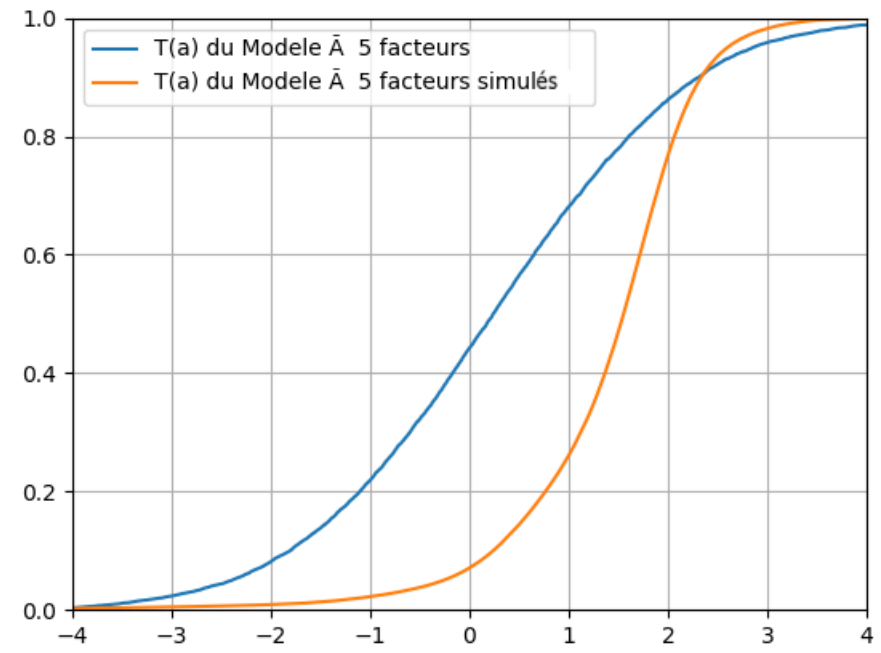

Figure 31: Cumulative density function for 5 factors model 2005-2019 Group 1000 millions AUM